\documentclass{aa}
\usepackage{graphicx, epsfig, fancyhdr, rotating, amsmath, epsf, txfonts, natbib, epstopdf, multirow}
\usepackage{subfigure, tabularx, appendix, dcolumn, threeparttable, longtable, makecell}
\usepackage{psfig}

\newcommand{\EQ}{\begin{equation}}
\newcommand{\EN}{\end{equation}}
\newcommand{\EQA}{\begin{eqnarray}}
\newcommand{\ENA}{\end{eqnarray}}
\newcommand{\Table}[1]{Table~(\ref{#1})}
\newcommand{\Sec}[1]{Section~\ref{#1}}

\begin{document}

\title{Eruptions from Quiet Sun Coronal Bright Points. II. Non-Potential Modeling}

\author{Klaus Galsgaard\inst{1}, Maria S. Madjarska\inst{2}, Duncan H. Mackay\inst{3}, Chaozhou Mou\inst{4}\thanks{The name of this author was wrongly spelled in Mou et al. (2018, Paper~I) as Chauzhou Mou.}}

\institute{
Niels Bohr Institute, Geological Museum, {\O}stervoldgade 5-7, 1350 Copenhagen K, Denmark
\and 
Max Planck Institute for Solar System Research, Justus-von-Liebig-Weg 3, 37077, G\"ottingen, Germany
\and 
School of Mathematics and Statistics, University of St Andrews, North Haugh, St Andrews, KY16 9SS, Scotland, UK
\and 
Shandong Provincial Key Laboratory of Optical Astronomy and Solar-Terrestrial Environment, Institute of Space Sciences, Shandong University, Weihai, 264209 Shandong, China
}

\date{Received date, accepted date}

\abstract
{Our recent observational study shows that the majority of coronal bright points (CBPs) in the quiet Sun are sources of one or more eruptions during their lifetime.}
{Here, we investigate the non-potential time dependent structure of the magnetic field of the CBP regions with special emphasis on the time evolving magnetic structure at the spatial locations where the eruptions are initiated.}
{The magnetic structure is evolved in time using a Non-Linear Force Free Field (NLFFF) relaxation approach, based on a time series of Helioseismic and Magnetic Imager (HMI) longitudinal magnetograms. This results in a continuous time series of NLFFFs. The time series is initiated with a potential field extrapolation based on a magnetogram taken well before the time of the  eruptions. This initial field is then evolved in time in response to the observed changes in the magnetic field distribution at the photosphere. The local and global magnetic field structures from the time series of NLFFF field solutions are analysed in the vicinity of the eruption sites at the approximate times of the eruptions.}
{The analysis shows that many of the CBP eruptions reported in \citet{2018A&A...619A..55M} contain twisted flux tube located at the sites of eruptions. The presence of flux ropes at these locations provides in many cases a direct link between the magnetic field structure, their eruption and the observation of mini coronal mass ejections (mini-CMEs). It is found that all repetitive eruptions are homologous.}
{The NLFFF simulations show that twisted magnetic field structures are created at the locations hosting eruptions in CBPs. These twisted structures are produced by footpoint motions imposed by changes in the photospheric magnetic field observations. The true nature of the micro-flares remains unknown. Further 3D data-driven MHD modelling is required to show how these twisted regions become unstable and erupt.}


\titlerunning{BP eruptions, modeling}
\authorrunning{K. Galsgaard et al.}

\maketitle

\section{Introduction}
\label{sect:intro}

The solar corona consists of a large selection of different phenomena that manifest themselves in X-ray and extreme-ultraviolet (EUV) observations over a variety of length and time scales. Generally, the main focus of interest is on the large-scale active regions (ARs) with all their complexity and highly time-dependent evolution. This, however, has a clear disadvantage. The high complexity of ARs is significantly challenging since their dynamical evolution is influenced by many different phenomena that act over a variety of length and time scales. At the other end of the size spectrum, the so-called coronal bright points (CBPs) represent one of the most typical omnipresent small scale phenomenon in the solar corona. They are found in the quiet Sun, coronal holes and in the vicinity of ARs. CBP plasma properties indicate that they represent a scaled down version of ARs \citep[for more details see][submitted]{madj_lrsp}. CBPs have a much simpler magnetic structure ($\le$60\arcsec\ in diameter) and shorter life spans (in EUV -- $\leq$20~hrs, $\leq$12~hrs in X-rays) that permits us to follow their full lifecycle \citep[e.g.,][]{1974ApJ...189L..93G, 1993AdSpR..13...27H,2018A&A...619A..55M}. These  properties of CBPs provide a unique opportunity to reach a better understanding of the basic physical mechanisms of coronal heating and dynamics.

The first paper of this study \citep[][hereafter Paper~I]{2018A&A...619A..55M} showed that more than two third (31 out of 42 or 76\%) of CBPs host at least one eruption during their lifetime. The study explored the observational properties of 11 quiet Sun CBPs and 21 eruptions associated with them. These eruptions took place $\sim$17~hrs after the CBP formation where the average lifetime of the CBPs in data taken in the 193~\AA\  channel of  the Atmospheric Imaging Assembly (AIA) on board the Solar Dynamics Observatory (SDO) was $\sim$21~hrs. They occurred during the convergence and cancellation phase of the CBPs' bipole evolution. The CBP eruptions presented an expulsion of chromospheric material either as an elongated filamentary structure (mini-filament, MF) or a volume of cool material (cool plasma cloud, CPC). This was accompanied by the ejection of the CBP or higher overlying hot loops. In some cases, coronal waves were also identified. A micro-brightening called micro-flaring (McF) was observed in all erupting MFs/CPCs and was associated with the polarity inversion line (PIL) of the bipoles related to the eruptions (only one case was inconclusive). The true nature of the McFs remains unknown. Mini coronal mass ejections (mini-CMEs) appear to take place in 11 out of the 21 CBP eruptions. Dimmings linked to the propagating mini-CMEs are seen as both `dark' cool plasma and areas of decreased coronal emission resulting from a plasma density depletion. This indicates that possibly mini-CMEs represent a characteristic part of the general CBP lifecycle, and that it is a natural stage in the evolution of CBPs. From Stereo EUVI data, \citet{2009A&A...495..319I} estimated that 1400 mini-CMEs (not specifically related to CBPs) happen per day on the whole Sun, while \citet{2012ApJ...746...12A} identified 2064 mini eruptions in AIA data. \citet{2018A&A...619A..55M} predicted that at least 870 eruptions in CBPs should occur if 76\% of all CBPs produce at least one eruption during their lifetime. 

A characteristic feature for the CBPs reported in Paper~I is the repeated occurrence of eruptions that are all related to magnetic flux convergence and cancellation in the photosphere. It has often been stated that convergence, combined with photospheric magnetic flux annihilation, is the main source for eruptions \citep[e.g. typically in the cases of classic CMEs,][]{2011LRSP....8....1C}. Various models of eruptions will be discussed in the next section. \citet{1986NASCP2442..369H} first reported a MF formation (average size 15\arcsec) along the PIL of small cancelling magnetic bipoles in the quiet Sun followed by the MF eruption. The process of mini-eruptions has not been investigated theoretically, although different cartoon models of mini-eruptions in association with blowout jets have been discussed on a number of occasions \citep[see the discussion in ][and the references therein]{2018A&A...619A..55M}. Jet models have been numerically investigated in relation to simple flux emergence scenarios, \citep[e.g.,][]{2013ApJ...771...20M, 2013ApJ...769L..21A}. These investigations show how the late phase of a flux emergence process may result in situations where eruptions can take place in different parts of the emerged bipolar region. 

Solar atmospheric observations represent 2D imaging of a 3D structure and therefore do not provide an easy way to determine the magnetic field structure of the observed phenomena. To obtain this structure some type of extrapolation of the magnetic field based on the underlying magnetogram is required. From such modelling, which is usually carried out using the potential field approximation, the structure of CBPs is often found to be a simple loop system connecting two well defined magnetic flux concentrations. A problem with this is that there is no simple way to provide free magnetic energy into these simple loop systems within their typical lifetime, and therefore it is not possible to explain why eruptions would take place in these systems at a given time during their lifetime. To investigate this issue the magnetic field and skeleton evolution during the lifetime of a CBP or at least for the certain period of time leading to the eruption time-period is required. 

The aim of this paper is to obtain a better understanding of the underlying magnetic field configuration that relates to eruptions in CBPs and to investigate how data-driven non-linear force free (NLFFF) modeling may help us to obtain an understanding of the dynamical build-up of the flux regions that host these eruptions. The article is organised in the following way. In \Sec{sect:model} a short review of eruption models is given. In \Sec{sect:obs} we describe the observations used for the time dependent evolution of the magnetic field, while the method applied is discussed in \Sec{sect:method}. \Sec{sect:res} discusses the findings for  a selection of  CBP eruptions studied here. For reference the full data sample is given in Appendix~\ref{details}. The results are discussed in \Sec{sect:disc}. Finally, the conclusions are given in \Sec{sect:sum}.

\section{Models of Eruptions}
\label{sect:model}
Eruptive phenomena in the solar corona have been discussed for many years, with many different models that describe the general magnetic field structure and its possible dynamical evolution. Many of these models are presented by cartoons, that depict the assumed structure of the magnetic field. These models are based on the observed 2D emission patterns and the 2D magnetic field distribution in the photosphere. Some models have subsequently been investigated in more detail, using both analytical approaches and numerical modelling. In the following we discuss some of these models.

When it comes to CME models, the characteristic model starts with a sheared arcade system, that experiences convergence of opposite polarity flux towards the PIL. \citet{1989ApJ...343..971V} using a cartoon model discuss the formation of prominences by shearing and converging of magnetic flux across the PIL. The model assumes that magnetic reconnection is responsible for the build-up of the flux rope (FR) that forms the prominence structure. A numerical attempt to make a twisted flux tube was made by \citet{1994ApJ...420L..41A} that sheared substantially an initial bipolar field and created a flux rope, showing that the basic idea works. This study has been extended by a series of investigations, where  \citet{2003ApJ...585.1073A, 2007ApJ...671L.189A, 2018Natur.554..211A} expand on the previous model by including a rotational twist of both flux concentrations, followed by a convergence phase. In \cite{2018Natur.554..211A} two models are compared, namely the flux cancellations model \citep{2000ApJ...529L..49A} against the breakout model \citep{1999ApJ...510..485A} and it was found that both models may work. The study further investigates the effect of the ambient magnetic field structure on the eruption and finally shows how this process works on a specific active region.
In a different study, an initial potential quadrupolar magnetic field is stressed by pushing two asymmetric footpoints into contact \citep{1999ApJ...510..444G}. This convergence provides the basis for a flux rope due to reconnection taking place in the lower atmosphere across the PIL. Finally, as an alternative process one could have a twisted flux system emerging into the corona \citep{2012NatPh...8..845R}. These models indicate that there exists several ways to build up an initial flux rope depending on both the prehistory and the path of the time dependent stressing. The exact process that takes place is determined by the evolution of the magnetic field configuration at the time close to the instability. 

Having created a flux rope, the eruption is a manifestation of an instability that drives the flux rope to expand upwards. Exactly what initiates this process is unclear. Theoretical explanations have been put forward \citep{2003A&A...406.1043T, 2005ApJ...630L..97T, 2010ApJ...718.1388D} describing different situations where a twisted magnetic field structure may experience an instability. From these it is clear that the flux rope requires a strong and highly twisted core, and then different possibilities exist. \citet{2003A&A...406.1043T}, \citet{2004A&A...413L..27T} and \citet{2005ApJ...630L..97T} investigated different scenarios where a twisted loop system becomes unstable and erupts. All cases contain a weak overlying magnetic field, which is unable to restrict the instability to a limited area. 
\section{Observations} 
\label{sect:obs}
The present study is a follow-up of the work in Paper~I where 11 CBPs were analysed in detail for their dynamical evolution as seen in several AIA \citep{2012SoPh..275...17L} channels. These cases were specifically selected for their position close to the disk center from the 70 cases identified in \citet{2016ApJ...818....9M}. Here we use the longitudinal magnetograms from the Helioseismic and Magnetic Imager\, \citep[HMI,][]{2012SoPh..275..207S} on board the SDO\, \citep{sdo2010} spacecraft  taken at 45\,s cadence and a 0.5\arcsec\ $\times$ 0.5\arcsec\ pixel size. The magnetic data were corrected for the projection effect in the image plane away from disk centre. For the modeling discussed below only 1 in every 10 images have been used, starting the selection from the beginning of the observed time series (see Paper~I for the time information). This provides a time resolution of 450\,s. The field-of-view of the analysed data is 512$\times$512 HMI pixels centred on the CBPs.  It would have been preferable for the modeling part of the analysis discussed below to use vector magnetograms, however, as the CBPs represent weak magnetic field regions, vector data is not available for the investigation \citep[for more details on the HMI noise level see][]{2009SoPh..260...83L,2014SoPh..289.3483H}.

\section{Numerical Method}
\label{sect:method}
The data preparation discussed in the previous section produces a long time series of magnetogram data representing a selection of CBP observations each of which covers a significant part of their lifetime. The data show various  typical features occurring during the CBPs lifetime including emergence, coalescence, fragmentation and cancellation, where eruptions are seen from all of the selected CBPs. In this paper, the processed magnetogram time series from Paper~I are used to approximate the time evolution of the magnetic field in the photosphere. The decreased time resolution of the HMI time series to 450\,s has two purposes. First, it eliminates the high frequency noise that exists between two consecutive magnetograms (45\,s), and therefore allows for a more clean determination of the systematic time changes of the magnetic features between subsequent magnetograms. Second, this long timescale variation is vital for following the slow systematic changes in the photospheric magnetic field configuration that systematically stress the 3D coronal magnetic field configuration over time. 

To simulate the 3D coronal evolution of the CBP directly from the magnetogram observations a time dependent NLFFF relaxation approach as described by \citet{2014ApJ...782...71G} is adopted. This technique follows the near ideal time evolution of the 3D magnetic field structure of the CBP regions where the evolution of the coronal field is driven directly by the magnetogram data. This approach is described in more detail in the following paragraphs.

As a preparation for the time evolution, the 2D HMI magnetograms were passed through a pre-processing process that allows for various cleanup and smoothing procedures. The corrected HMI magnetograms represent the boundary conditions for both the initial potential 3D magnetic field and its subsequent time evolution using the NLFFF relaxation technique. Apart from the procedures applied to the magnetograms before their use (see Section~3), only single pixel clusters with unrealistic high count values have been removed, and no lower flux threshold for zeroing pixels values has been adopted. This combination was chosen as the flux regions considered in this investigation, are small in pixel sizes compared to active regions for which the code is normally used \citep{2014ApJ...782...71G, 2018ApJ...852...82Y}. In the present study removing flux below a given threshold may strongly influence the magnetic field topology and the derived time evolution. In the following the continuous time evolution of the magnetic field obtained through the applied magnetograms is assumed to be 2D periodic in the horizontal direction. This allows a Fast Fourier Transform (FFT) approach to derive the initial potential 3D magnetic field. As none of the magnetograms are in a perfect flux balance, the top boundary of the domain is open which allows the excess magnetic flux to exit. 

The NLFFF relaxation code uses the magnetic vector potential, $\bf{A}$, as the primary variable. To simulate each CBP a time series of vector potentials are derived at the photosphere, based on the observed $B_z$ magnetic field. To change the magnetic field on the photospheric boundary in accordance with the observations, it is assumed that the two horizontal components of the vector potential in the photospheric plane can be represented by a scalar potential ($\Phi$) in the following way,
\EQA
A_x &=&  {{\partial \Phi} \over {\partial y}},\\
A_y &=& -{{\partial \Phi} \over {\partial x}}.
\ENA
Using the general definition of the magnetic field by a vector potential, $\bf{B} = \nabla \times \bf{A}$, and setting the gauge to zero, these two approaches are combined to provide a Poisson equation for determining the scalar potential $\Phi$ based on the knowledge of the magnetic field at the bottom boundary,
\EQ
{{\partial^2 \Phi} \over {\partial x^2}} + {{\partial^2 \Phi} \over {\partial y^2}} = - B_z.
\EN
Assuming the data in the 2D plane are periodic, this equation is solved using a FFT approach. This solution may be expanded in 
height defining an initial potential magnetic field using the Devore Gauge \citep{2000ApJ...539..944D}.

To simulate the evolution of the photospheric and coronal magnetic fields through a continuous sequence of NLFFF  solutions driven by the evolution of the photospheric field as deduced from the magnetograms, the following technique is applied. To start the simulation the vector potential {\bf$A$} describing the initial potential field is taken along with its deduced coronal field. Next the vector potential components at the base $(A_x, A_y)$ are updated such that it results in the time evolution
of the radial magnetic field at the photosphere from the present observed magnetogram to the next. The effect of this boundary evolution is to inject electric currents and non-potentiality into the coronal field which evolves the coronal field away from equilibrium. In response to this the vector potential in the full 3D domain is found by solving the uncurled induction equation, 
\EQ
{{\partial {\bf A}} \over {\partial t}} = {\bf v} \times {\bf B} + {\bf R_{num}}
\EN
where ${\bf v}$ is the magneto-frictional velocity, expressed by
\EQ
{\bf v} = {1 \over \nu} {{{\bf j} \times {\bf B}} \over B^2},
\EN
and ${\bf R_{num}}$ is a non ideal term that allows for numerical diffusion. The role of the magneto-frictional velocity is to return the coronal field to an equilibrium force-free state, in general a non-linear force-free field. Using this technique a continuous time sequence of NLFFF can be produced from the observed magnetograms. A full description of the code is given in \citet{2011ApJ...729...97M} and \citet{2014ApJ...782...71G}.

For each update of the boundary conditions, provided by the selected HMI data, the induction equation is solved in a frictional time until the magneto-frictional velocity becomes sufficiently low. This indicates that a new near NLFFF state has been reached and a snapshot of the 3D vector potential, $\bf A$, is saved. From each HMI data-set the 3D relaxed magnetic field is determined. This 3D vector field is analysed, using {\it Vapor}, in an attempt to better understand the structural evolution of the magnetic field, with an emphasis on the region around the erupting CBP. 

\section{Results}
\label{sect:res}
Using the technique described in \Sec{sect:method}, analysis of the morphological and dynamical time evolution has been conducted for the 11 CBPs discussed in Paper~I. In the following the 3D magnetic field structures that hosts the eruptions deduced from the time-dependent NLFFF relaxation approach described above are discussed. The analysis is based on a comparison between the observations shown in Paper~I, e.g. Figures~1 and 3, and the modelled magnetic field structure. The figures and movies in Paper~I combine information from HMI and 4 different AIA channels (304~\AA, 171~\AA, 193~\AA\ and 94~\AA) to indicate the morphology (including specific features) and dynamics of the eruptions as well as their timing and location with respect to the CBP evolution.

A discussion of a limited number of the eruptions is presented here. The discussion of the remaining cases can be found in Appendix~\ref{details} below. We also provide animations of the models using a 450~s  cadence that cover a time period  of $+/-$75 minutes on either side of the observed time of eruption. The starting points of the magnetic field line traces are chosen to indicate the field line structure at the time of eruption and are fixed in time. Therefore, a small number of field lines are seen to change between closed and open status between successive frames. The field lines are colour coded by the amount of current.

\Table{tab1.ref} contains the general parameters and information of the different eruptions in relation to the NLFFF relaxation process. For more information on the eruptions consult Table~(2) in Paper~I and the general discussion of the CBPs there.

\begin{table*}
\centering
\caption{General information on the CBP eruptions and the configuration of the associated coronal magnetic field.}
\label{tab1.ref}
\begin{tabular}{ccccc}
\hline
\hline
No. & Start time for eruption & Duration & Local Field line structure & Global field line structure \\
    & data end time           & minutes  &                            &                             \\ 
\hline
1 & 02/01/2011 22:31 & 25 & Small-scale Arcade System (SsAS) & Fan-spine structure (FS) \\
  & 03/01/2011 04:42 & 24 & SsAS with a Flux Rope (FR) & FS \\
\hline
2 & 01/01/2011 19:25 & 20 & SsAS with a FR & FS \\
  & 01/01/2011 23:21 & 16 & SsAS with a FR & FS \\
\hline
3 & 02/01/2011 09:23 & 17 & SsAS with an external FR & Large Arcade System (LAS) \\
  & 02/01/2011 13:12 & 33 & SsAS with an external FR & LAS \\
\hline
4 & 02/01/2011 03:32 & 20 & SsAS with complications and  FR        & LAS containing several footpoints \\
  & 02/01/2011 05:18 & 31 & SsAS with complications and  FR        & LAS containing several footpoints \\
  & 02/01/2011 06:24 & 19 & FR region with one polarity footpoints & Complex overlying loop system \\
\hline
5 & 01/01/2011 21:36 & 15 & SsAS with an external FR & Open Field (OF) and LAS \\
\hline
6 & 03/01/2011 10:04 & 19 & SsAS with a FR & LAS with different connectivities \\
\hline
7 & 01/01/2011 00:31 & 17 & SsAS               & LAS \\
  & 01/01/2011 15:31 & 13 & SsAS/LAS with a FR & LAS \\
  & 01/01/2011 16:17 & 33 & SsAS/LAS           & LAS \\
  & 01/01/2011 18:42 & 18 & SsAS with a FR     & LAS with a FR \\
\hline
8 & 02/01/2011 21:10 & 40 & SsAS with a FR & OF and LAS \\
  & 02/01/2011 23:42 & 22 & SsAS with a FR & OF and LAS \\
\hline
9 & 03/01/2011 13:30 & 35 & SsAS with a FR from one polarity footpoint & Complex field structure\\
\hline
10 & 03/01/2011 13:53 & 22 & SsAS with different connectivity & Complex null structure \\
\hline
11 & 02/01/2011 09:10 & 47 & SsAS with a FR & LAS \\
   & 02/01/2011 22:10 & 40 & SsAS with a FR & LAS \\
\hline
\end{tabular}
\end{table*}

The main focus of the present investigation is the local and global 3D magnetic field structures at the locations of the observed eruptions. The local 3D magnetic structure probes the connectivity between the local bipolar flux concentrations, while the global structure explores the connectivity with the nearby flux concentrations. We follow the notation from Paper~I, where {\it BP\#  ER\#} represents the eruption number for a given CBP. Movies for the observations can be found here: \url{https://doi.org/10.5281/zenodo.1300416} and for the modelling here: \url{https://doi.org/10.5281/zenodo.1915477}
 

\begin{figure*}
{\hfill
\includegraphics[scale=0.23]{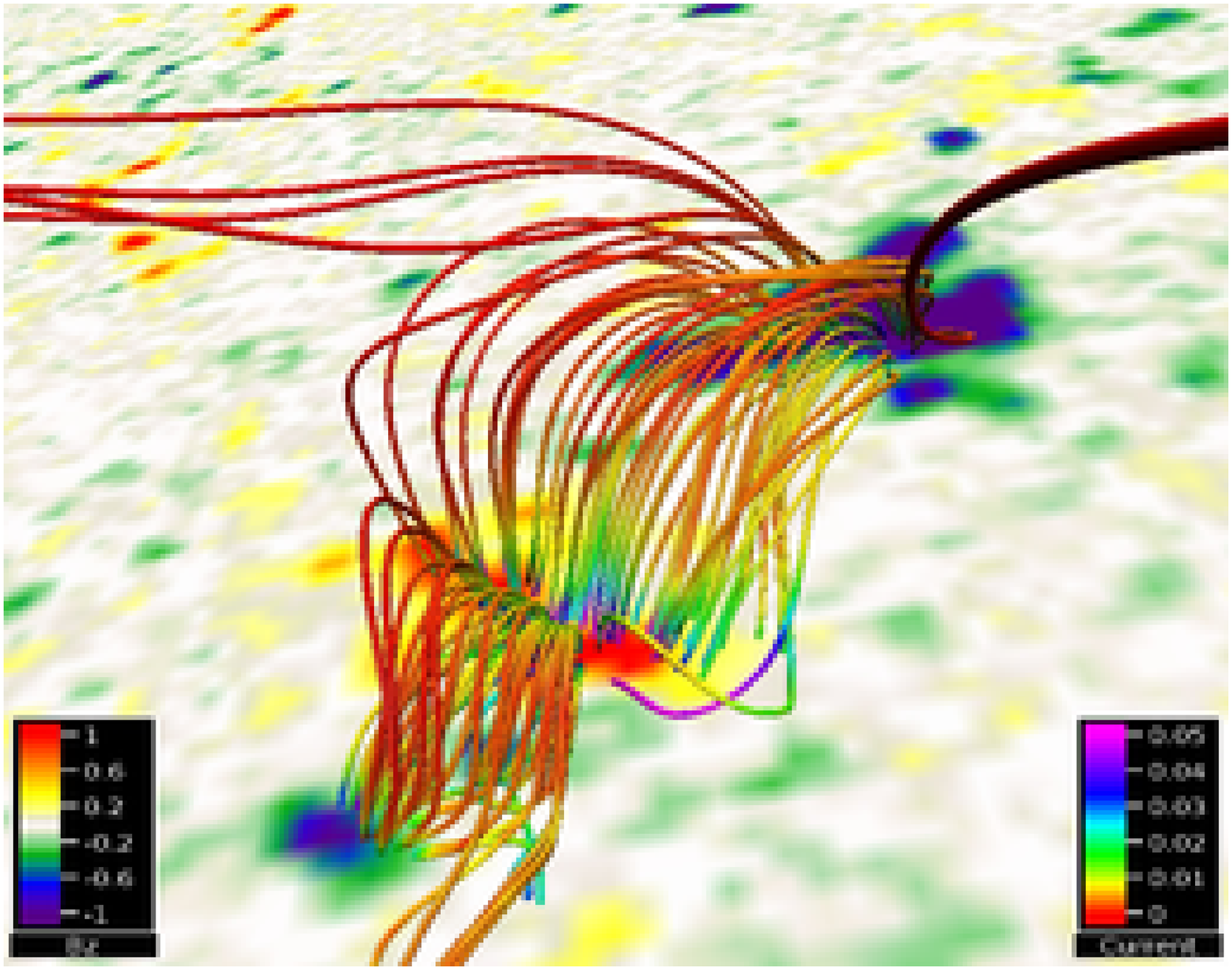}
\hfill
\includegraphics[scale=0.23]{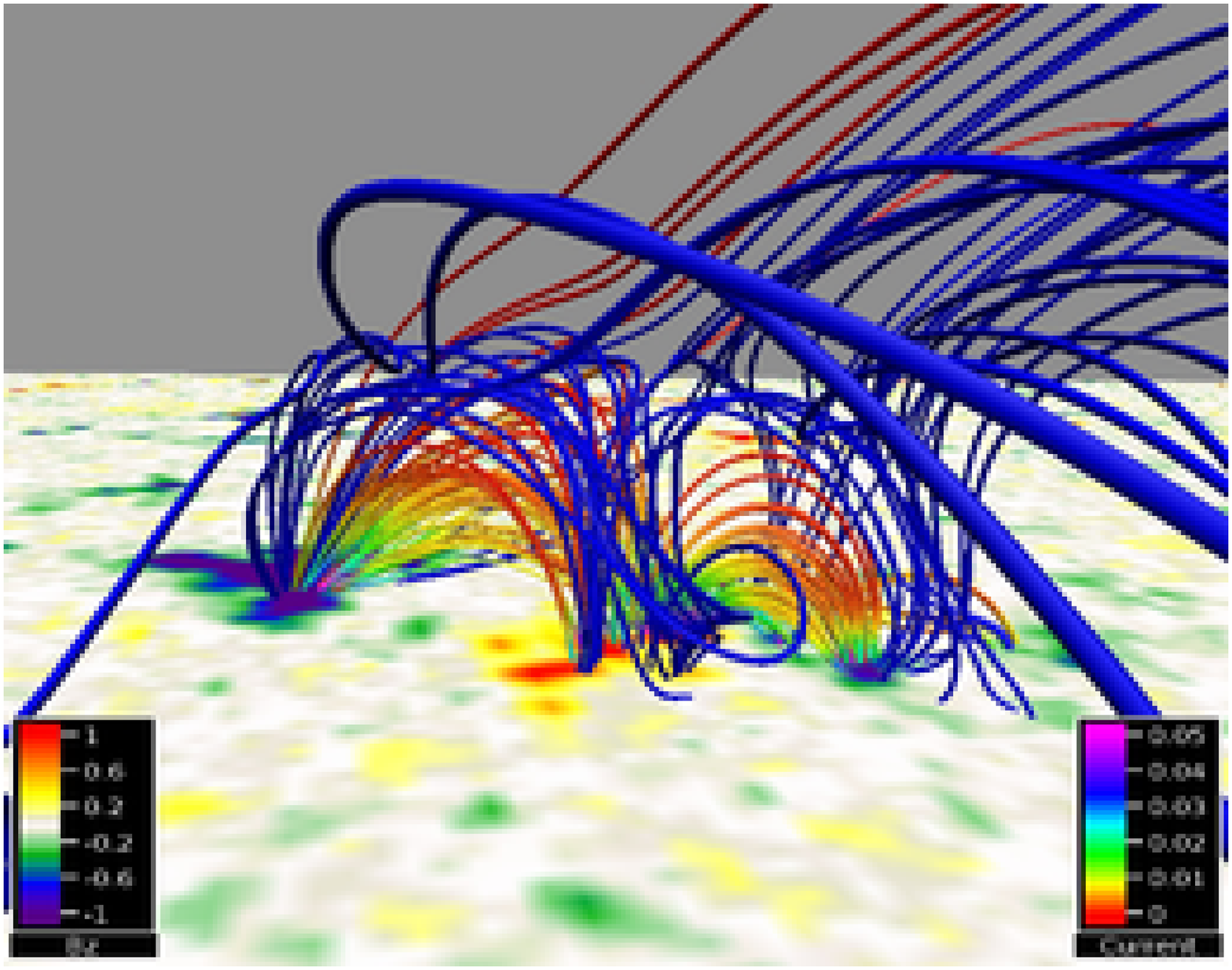}
\hfill}

\caption[]{{\it BP1--ER1} --  Left panel, the loop structure of the BP1 region and the neighbouring smaller loop system that host the first observed eruption. The bottom plane represent the magnetogram scaled to $\pm$ $85.3$ Gauss. The field lines are colour coded with a non-dimensional magnitude of the current. The current amplitude is in units of $0.0188$ Ampere. The right panel additionally includes the large scale magnetic field (blue field lines) that embeds the loop systems seen in the left panel. Time: 22:39 UT. An animation (BP1\_erupt1.mov) associated to the left panel is available in the electronic edition.}
\label{BP1_1}
\end{figure*}

{\it BP1--ER1:} From the time series of the NLFFF simulation data, the magnetic field structure is investigated. The left panel of Fig.~\ref{BP1_1} shows two arcade structures rising from the positive polarity of the CBP . The eruption takes place at the PIL of the southern loop system which is smaller (see the movie associated with Fig.~\ref{BP1_1}). We classify this as a {\it Small-scale Arcade System} (SsAS). This SsAS maintains its simple loop structure around the time of the eruption. There is no indication of a twisted magnetic field structure embedded in the loop structure that could produce the observed formation of the MF and its eruption. The size of the region where the eruption takes place is small and is only resolved with a small number of HMI pixels. When numerically stressing the magnetic field, the low pixel resolution of the region implies that numerical diffusion has a negative effect on the possible build-up of magnetic field line complexity. The erupting SsAS region does show an enhanced electric current concentration around the PIL, to indicate an enhanced amount of free magnetic energy for this location, where the micro-flare is observed. The right panel of Fig.~\ref{BP1_1} represents the large scale field line structure of the CBP and erupting loop system. 
 This shows how the two loop systems seen in the left panel of Fig.~\ref{BP1_1} are embedded in a loop like structure that connects to a remote positive flux concentration to the south-east of the CBP. This combined magnetic configuration is comparable to a `fan-spine' structure defined by a single 3D null point located above. However, to identify a null point in this specific case, the positive flux concentration should be completely  surrounded by the negative flux at the photospheric surface. The lack of a clear identification of a null point in the structure is due to the horizontal extent of the two loop systems.
In the observation the MF is seen moving in the south-east direction. The overlying field above the erupting MF loop system therefore seems to guide any upward moving structure in the same direction.


\begin{figure*}
{\hfill
\includegraphics[scale=0.23]{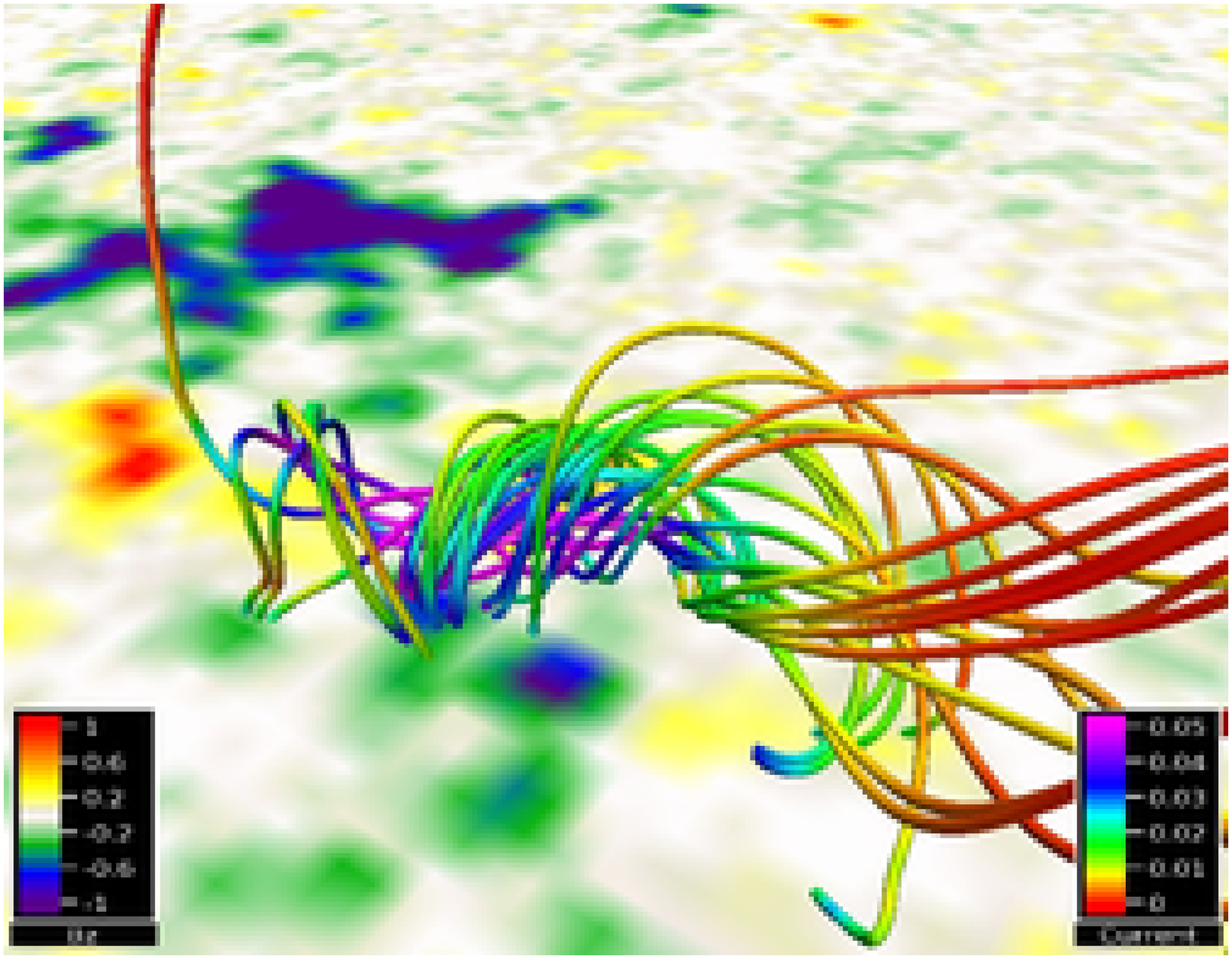}
\hfill
\includegraphics[scale=0.22]{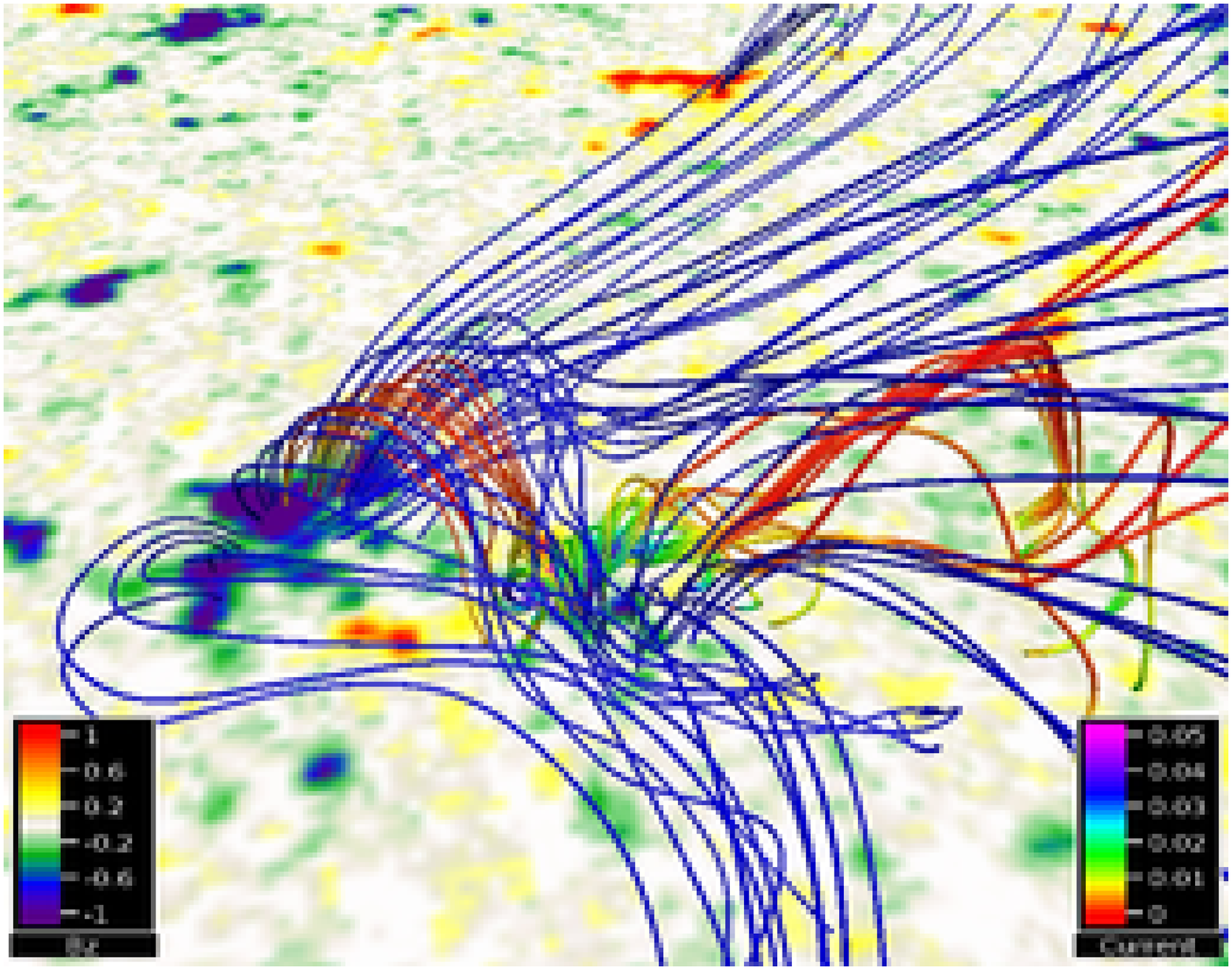}
\hfill}
\caption{As in Fig.~\ref{BP1_1} but for {\it BP1--ER2}. Left panel shows the field line structure at the region where the second eruption in BP1 takes place.  Time: 04:49 UT. An animation (BP1\_erupt2.mov) associated to the left panel is available in the electronic edition.} 
\label{BP1_2} 
\end{figure*}

{\it BP1--ER2:} The second eruption takes place at the south edge of the CBP region, which is the same area that produced the first eruption nearly 5 hours earlier. The eruptions can therefore be seen as homologous events (for the definition of homologous events and overview see Paper~I). The time evolution of the magnetic field in the erupting region shows the formation of a small magnetic flux rope (FR) that grows with time, even after the time of the eruption. The left panel of Fig.~\ref{BP1_2} shows the magnetic field structure at 04:49 UT. In comparison to the BP1--ER1 case, strong current is clearly seen on the low lying twisted field lines, indicating the presence of free magnetic energy. From the movie (see the movie associated with Fig.~\ref{BP1_2}) it is seen that twist and current increase with time until approximately the time of the eruption, when it then decreases. Further to this it is also found that most of the twisted field lines seem to be rooted in the negative flux concentrations only.  As a consequence they do not look like a traditional flux rope above a PIL. This finding is strongly supported by the MF statistical study  of \citet{1986NASCP2442..369H} who report that  one or both footpoints of the MF  were rooted at the bipole inversion line. The right panel of Fig.~\ref{BP1_2} shows both the local flux rope and the large scale field that has a Fan-spine (FS) structure. The FS structure seen in the first eruption is less obvious for this eruption. The main difference between the two eruptions is that the size of the loop system containing the  FR is clearly smaller in the second eruption. 


\begin{figure*}
{\hfill
\includegraphics[scale=0.23]{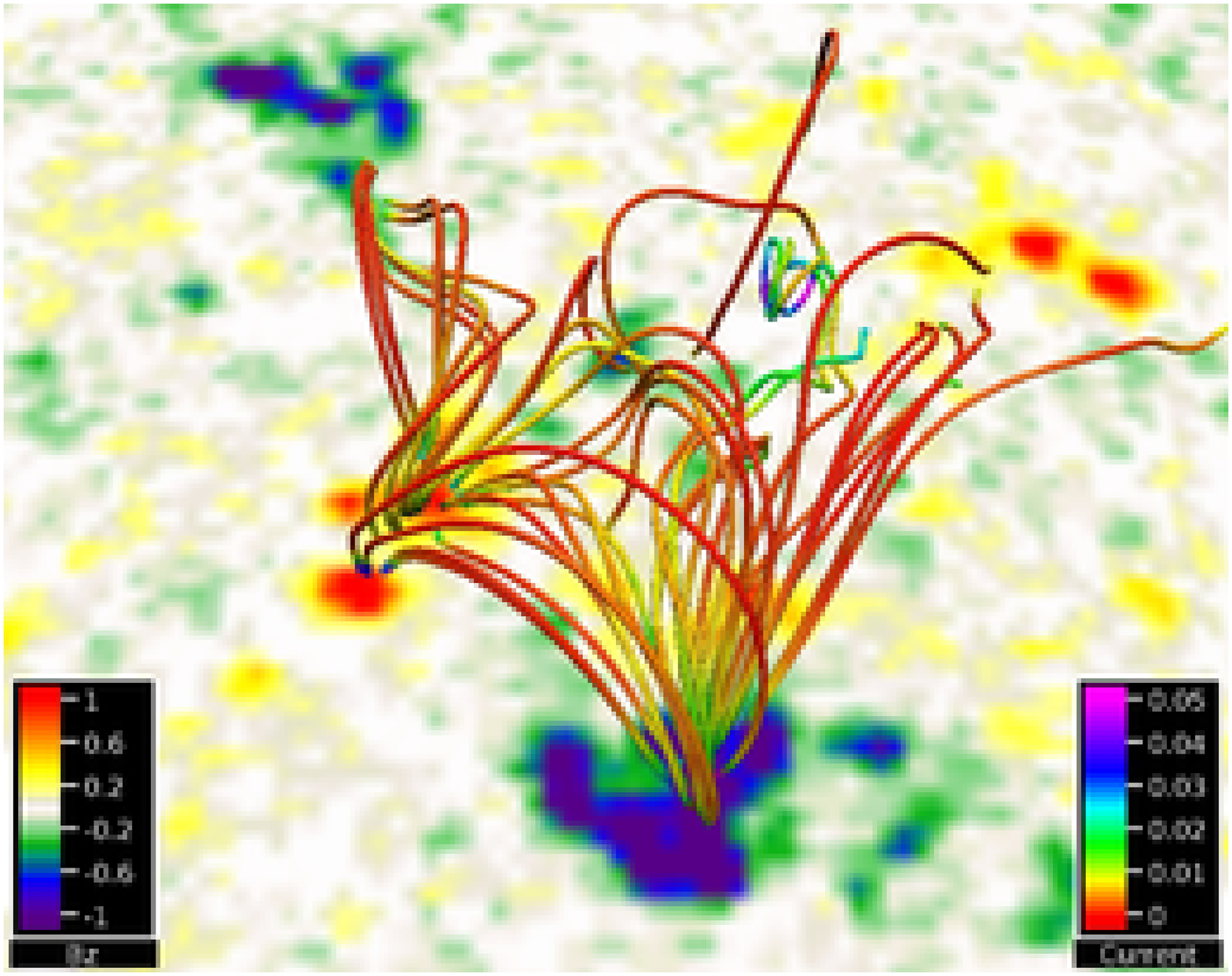}
\hfill
\includegraphics[scale=0.23]{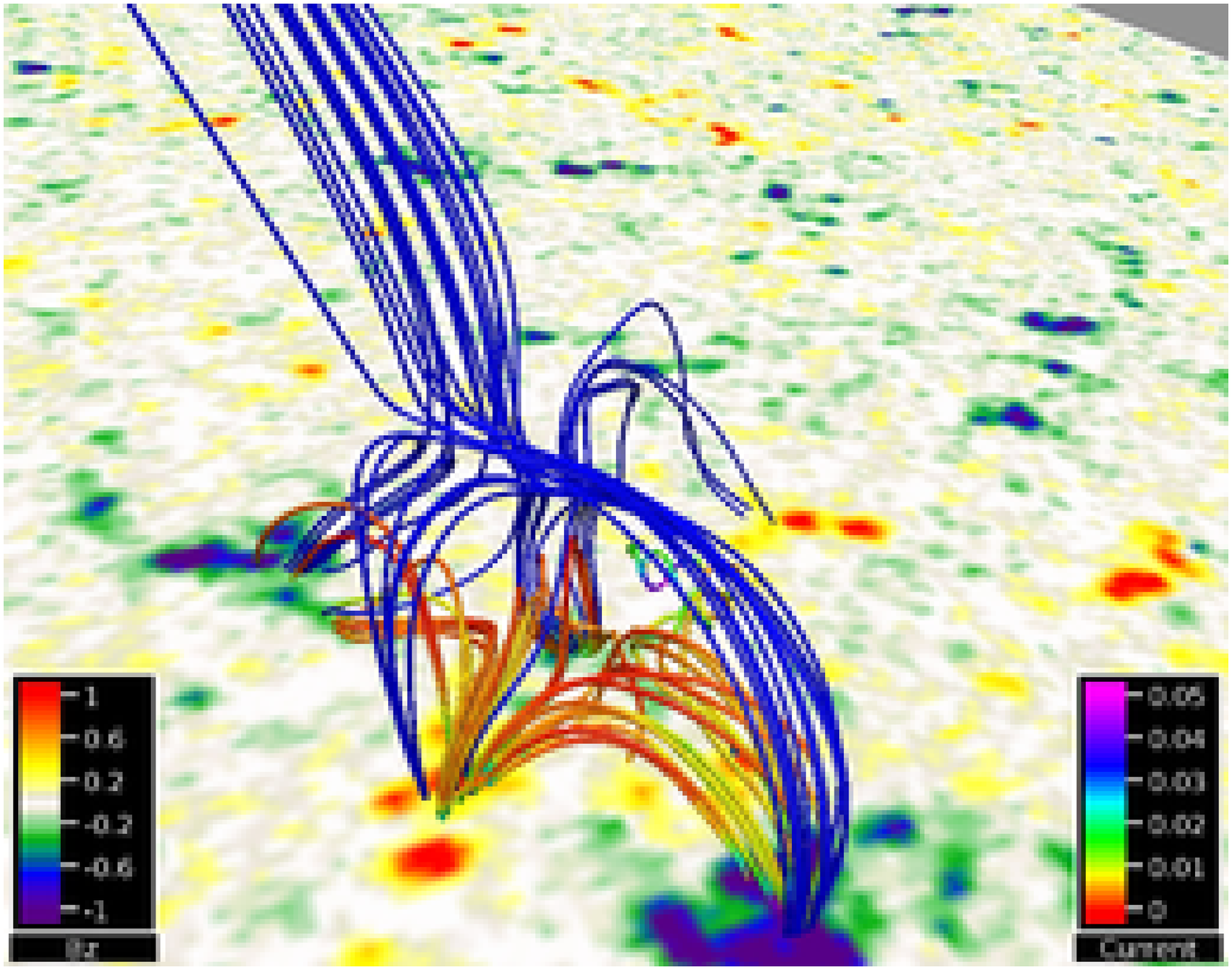}
\hfill}
\caption{As in Fig.~\ref{BP1_1} but for BP10. Notice the presence of the very skew 3D null point represented by the blue magnetic field lines. Time: 13:39 UT. An animation (BP10\_erupt1.mov) associated to the left panel is available in the electronic edition.}
\label{BP10_1} 
\end{figure*}

{\it BP10:}  The local field line structure of this CBP represents a series of SsAS, left panel of Fig.~\ref{BP10_1} (see the movie associated with Fig.~\ref{BP10_1}), which produces a complicated region where the CBP emission originates (see also Fig.~A.17 in Paper~I). The positive flux patch connects to several negative flux concentrations in the local vicinity. Around the location of the first emission, the magnetic field is divided in terms of connectivity, creating a location where the magnetic field contains a highly distorted 3D magnetic null point, right panel of Fig.~\ref{BP10_1}. This null point has the spine axes rooted in the negative polarities on either side of the positive polarity and the fan surface connects down to the local positive polarity patches  (A magnetic field structure that is different from the FS structure seen in BP1 and BP2). It is likely that this null point may be the key for the magnetic energy release process associated with this eruption. The location of the null point and the initial emission enhancement in the AIA images seems to be well aligned in space. The spreading of the dimming seems to follow the local division line of the fan surface from the null point towards the north-west and south-east. This indicates that the eruption takes place within this domain of the magnetic field and seems consistent with the energy release being controlled by the magnetic null point.


\begin{figure*}
{\hfill
\includegraphics[scale=0.23]{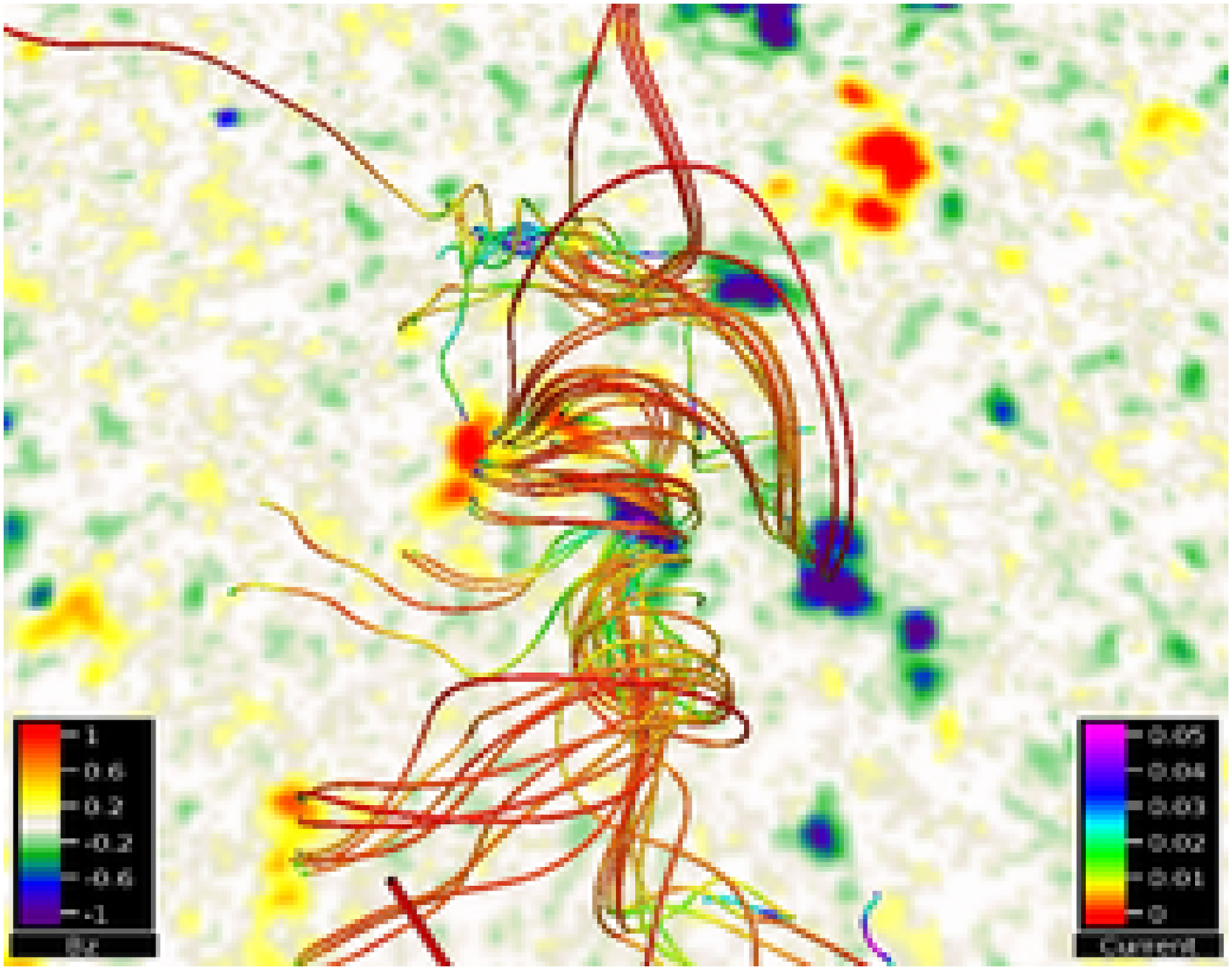}
\hfill
\includegraphics[scale=0.23]{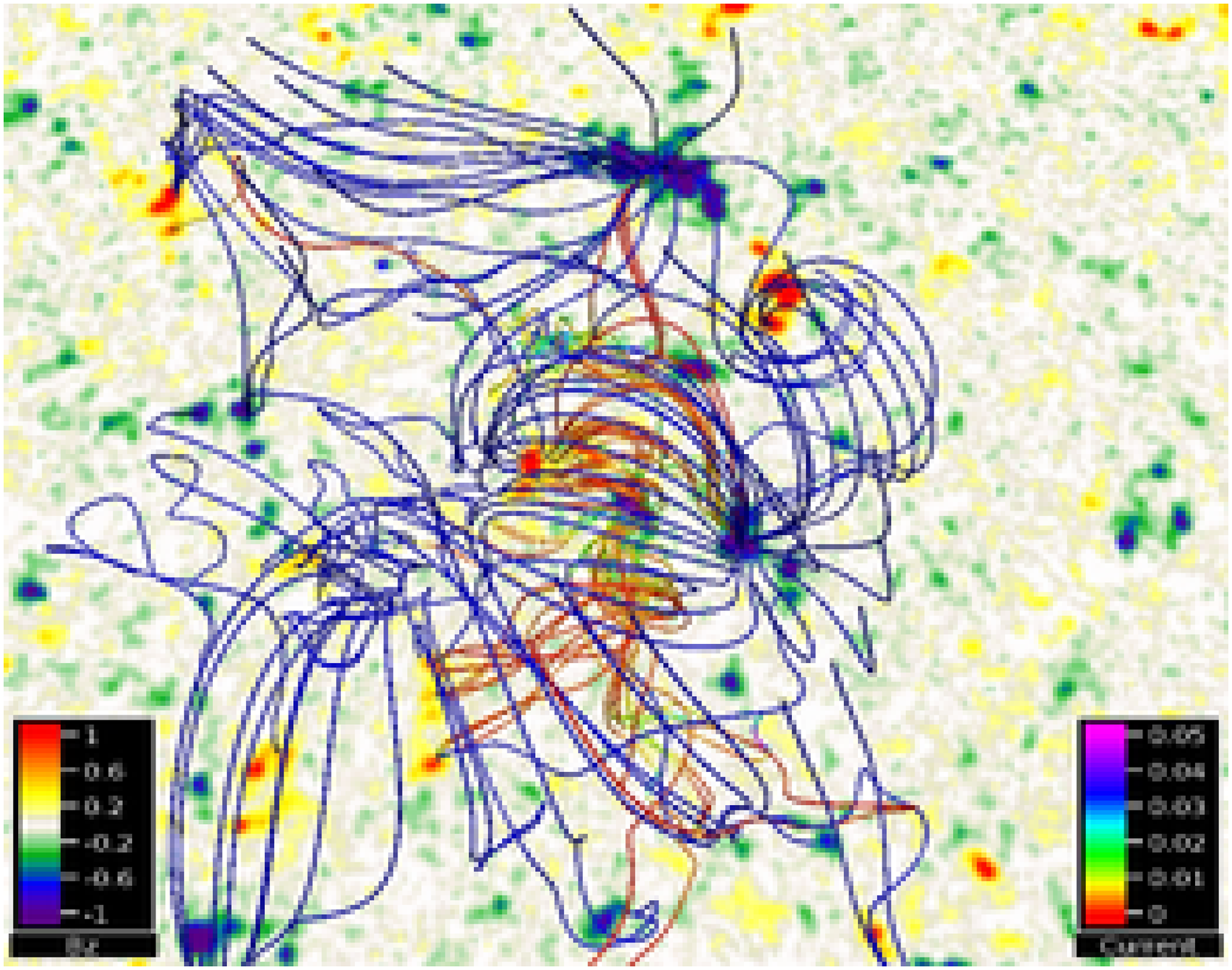}
\hfill}
\caption{As in Fig.~\ref{BP1_1} but for BP11--ER1. Time: 9:45 UT. An animation (BP11\_erupt1.mov) associated to the left panel is available in the electronic edition.}
\label{BP11_1}
\end{figure*}

{\it BP11--ER1:} The first eruption of BP11 takes place in a relatively simple bipolar region, left panel of Fig.~\ref{BP11_1} (see the movie associated with Fig.~\ref{BP11_1}). The field lines tracing from this region show a SsAS with very little twist connecting in the east-west direction around the bipolar region. Extending the region to the north and south of this region, twisted structures are found. Both of these FRs contain only a weak current. Looking at the field line structure close to the region of emission, the  FR to the south is very clear, while also the second FR to the north of the bipolar region is identified. How any of these FRs impact the eruption is speculative. The large scale overlying magnetic field represents large arcade system (LAS) that over a large length scale encapsulates the underlying loop system. Further away from the large scale structure the field connectivity seems very complex which does not provide a means to guide the expansion of the ejecta from the eruption. This LAS  could explain the two dimming regions that expand north and south of the CBP (see Fig.~4 in Paper~I) confining  the erupting cool material of the MF and the expanding CBP loops.


\begin{figure*}
{\hfill
\includegraphics[scale=0.23]{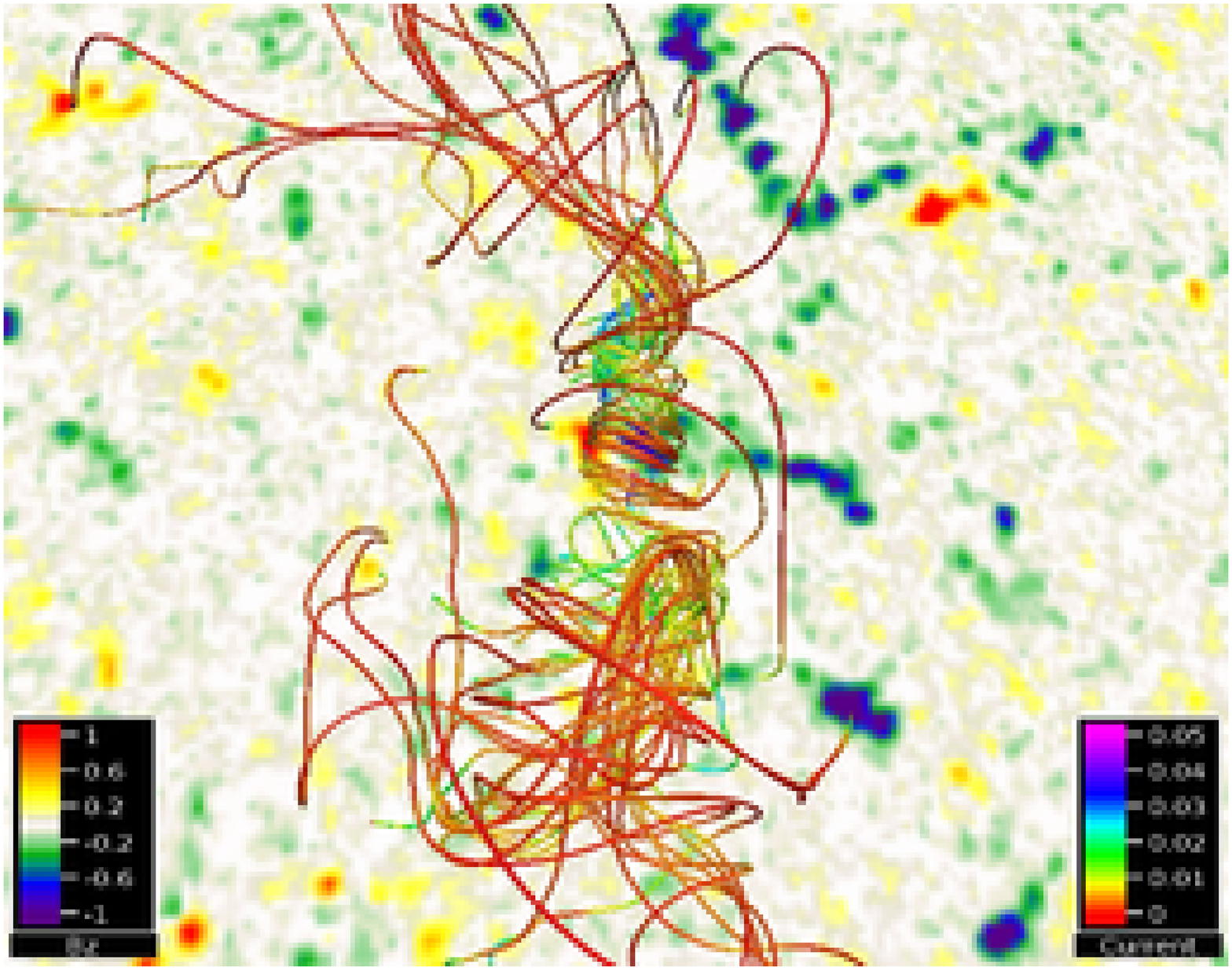}
\hfill
\includegraphics[scale=0.23]{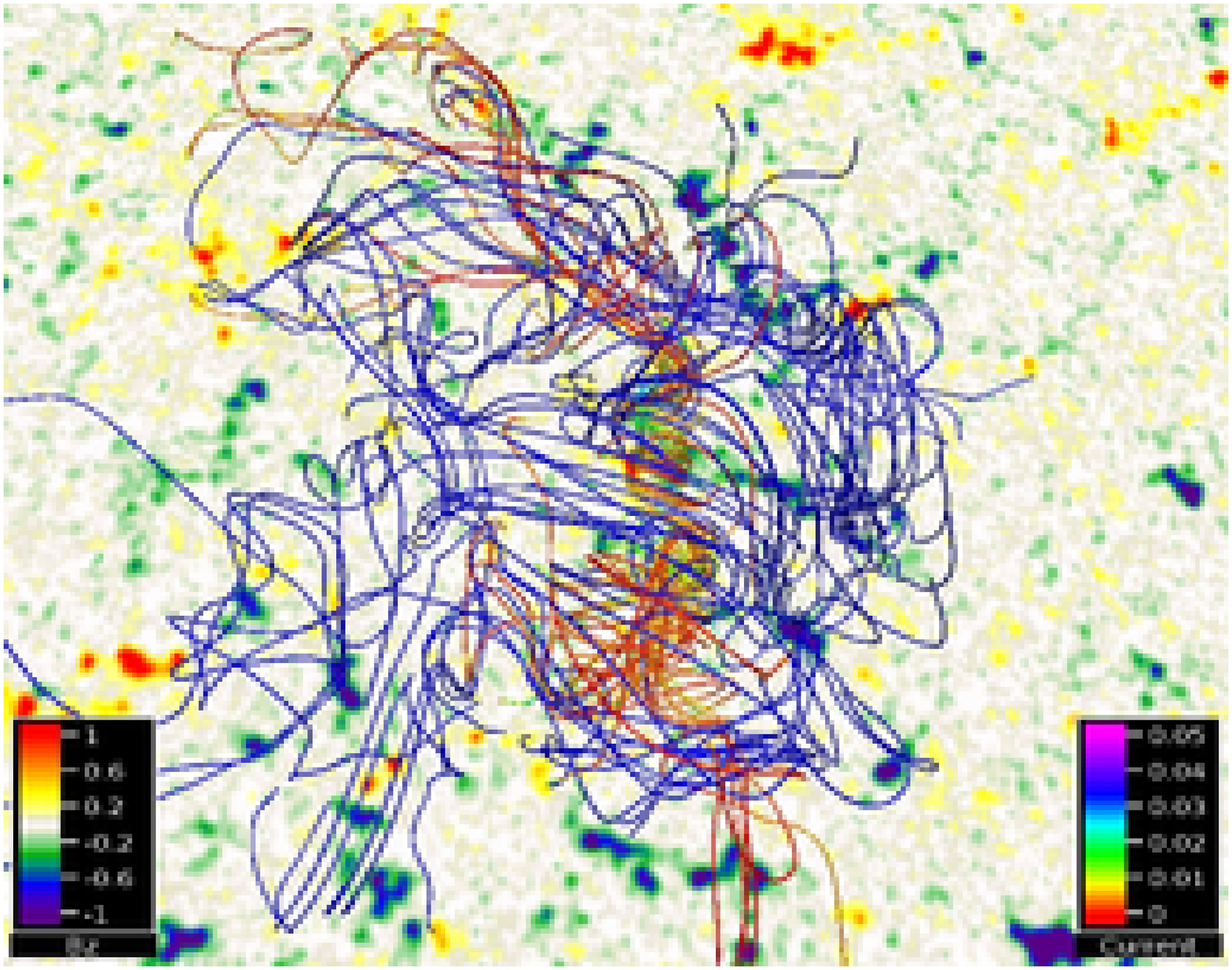}
\hfill}
\caption[]{As in Fig.~\ref{BP1_1} but for BP11--ER2.  Time: 22:04 UT. An animation (BP11\_erupt2.mov) associated to the left panel is available in the electronic edition.}
\label{BP11_2} 
\end{figure*}

{\it BP11--ER2:} The flux structure of this eruption is simpler than the first eruption, and the effect of the eruption is relatively small, as   also confirmed by the observations (see Fig.~6 in Paper~I). The field line structure of the region is shown in the left panel of Fig.~\ref{BP11_2} (see the movie associated with Fig.~\ref{BP11_2}). This shows that the field line structure in the CBP region is a SsAS, with a clear  FR extending to the north. The eruptive region to the north-east contains both sheared arcade structures between the positive and negative polarities, but also a more complicated  FR extending to the east of the loop system. The structure for the eruption to the south-east resembles the one seen in BP10 with the null point in the middle of the structure. It is therefore likely that three independent regions become unstable and erupt over a very short time period. The large scale magnetic field, right panel of Fig.~\ref{BP11_2}, shows a LAS above the central CBP region.  
\section{Discussion}
\label{sect:disc}

In previous investigations of CBPs, potential magnetic field extrapolations have been used to model the local magnetic field structure deduced from magnetograms \citep[i.e.][]{2008A&A...492..575P, 2011A&A...526A.134A, 2012ApJ...746...19Z, 2017A&A...606A..46G}. These investigations are often limited to a single snapshot in time and in size, where the constraint of the lowest energy solution only illustrates the bipolar nature of the CBP region.

To allow the time evolution of the model considered here to be more realistic than previous static investigations, a sizeable region around the CBPs has been used (512$\times$512 HMI pixels). This size of region is required to permit both the time evolution of the local small-scale magnetic field structure, and the large scale magnetic field structure of the CBP region to be investigated. The time evolution allows electric current carrying structures in the force free magnetic field to build up and create magnetic structures that may become unstable and erupt. The large scale modelling is required for investigating the expansion path of the eruption ejecta away from the initiation region. 

The numerical calculations presented here have been executed without an explicit diffusion parameter, in order to maintain, as much as possible, the field line connectivity due to the imposed boundary changes of the initial magnetic field configuration. As the applied code is not free from numerical diffusion, there is a lower limit in grid resolution to which more complicated magnetic structures can be assumed to build up. Below this length-scale numerical diffusion may be significant, and may therefore constantly be removing the tendency to build up for instance twisted flux structures. The amount of observed twist in structures close to the resolution limit may therefore be smaller that it would have been if the magnetic field evolution had been truly ideal. 

Even with this restriction in the simulation data, twisted flux structures are identified in the simulation domains. These are a result of the continuous photospheric motions of the magnetic footpoints combined with the process of flux cancellation. These motions `braid' the magnetic field in various ways and on a variety of different length and time scales. Within the NLFFF relaxation simulations, magnetic flux ropes may form due to two main reasons. The first is a consequence of flux cancellation in the photosphere between the footpoints of two sheared arcades \citep{1989ApJ...343..971V, 2014ApJ...782...71G}. This process may occur along a PIL as the vertical components of the field cancel but the horizontal component remains along the PIL. Multiple occurrences of this may then produce a flux rope. The second method is through numerical diffusion within the coronal volume as highly stressed fields are produced. Additionally rotational motions of flux concentrations may produce twist, but such motions are not easily identified nor imposed for the method used here to drive the simulation. This would be possible if vector magnetogram data 
covering the quiet Sun existed.

We note that the NLFFF relaxation method is not the appropriate way to follow the evolution through a real instability, where magnetic reconnection plays an important role in changing the magnetic field line connectivity over a short time scale. Previous studies on both the local and global scale have shown the occurrence of non-equilibrium within similar NLFFF simulations \citep{2006ApJ...641..577M, 2006ApJ...642.1193M, 2014ApJ...782...71G, 2018ApJ...852...82Y}. Such events are seen where a flux rope first forms and then increases in size. Eventually the flux rope may become too large to be held down by the overlying arcades and as a result it starts to rise. As it rises reconnection either through a specified non-ideal term or due to numerical diffusion may occur below the flux rope. This then produces an enhanced outward radial Lorentz force that forces the flux rope to rise and subsequently be ejected out of the computational box if open boundary conditions are used. After the flux rope is expelled the system returns to a new equilibrium state. We note however that the ejection of the flux rope does not occur over a true physical timescale, and to capture the full dynamics a full MHD simulation is required \citep{2013A&A...560A..38P, 2018JSWSC...8A..26P}. It does however illustrate that the magneto-frictional approach may be used to follow the quasi-static evolution of the field as long as there is sufficient time between the ejections for the field to be able to eject the flux rope and return to a new equilibrium state.

In the cases discussed above, no eruptions are seen in the NLFFF simulations. The most likely reason for this is that the twisted flux regions that may become unstable are to close to the resolution limit to allow them to evolve into an unstable state. Local diffusion then becomes important for cleaning up the field structure before the field begins to build up new stress for the next eruption.   

Seven of the eleven CBPs discussed above have repeated eruptions from the same topological region of the magnetic field. In some of the cases the twist is seen to `diffuse' away before a new twisted structure builds up at nearly the same location providing the basis for the next eruption. In other cases, the twist is not removed between subsequent eruptions. In these cases the magnetic field structure is continually `braided'. 

A different investigation \citep{2013ApJ...771...20M} found that the emergence of magnetic flux into an open magnetic field, produced in the late phase of the emergence process a sequence of eruptions, with repeated eruptions taking place in the same volume of the emerged structure. Investigating these eruptions in more detail, it is found that the events contain a tether cutting process below the rising magnetic flux structure (a mini-CME like structure) and that this process adds twist to both the magnetic structure located below and above the current sheet. This increase in twist is important for the underlying flux structure to start its path towards a new eruption. The lack of cleaning up after the eruption in the relaxation model may therefore be less of a real problem than first expected, but more investigations of these phenomena are clearly needed.

The NLFFF relaxation model provides a possible technique for investigating the general structure of the magnetic field where repeated eruptions occur. It is found that these events can take place in the same magnetic field configuration for the same CBP. The different cases show that the global magnetic field structure of the eruptions are different between the different CBPs.  In BP1 and BP2 a similar FS like structure is present, while the structure in the other cases is less coherent in nature, with one case containing a different orientated 3D magnetic null point above the CBP region (BP10). In addition, 3D magnetic null points, having a FS structure, were also found to dominate the magnetic structure for a sample of CBPs located in coronal holes \citep{2017A&A...606A..46G}. An important question is how the large scale structure of the magnetic field influences both the local presence of the CBP and also the path of the ejecta from the eruptions. The various structures of the large scale magnetic field found in this investigation indicate that for the quiet Sun, relaxed magnetic field configurations have different magnetic configurations than those found in coronal holes \citep{2017A&A...606A..46G}. One important difference in the present study is the continued stressing of the magnetic field, that may alter the structure of the magnetic field, due to the allowed free magnetic energy. Future modeling studies could consider a comparison of potential field extrapolations with the NLFFF model. In the BP1 and BP2 cases the FS like structure seems to be an obvious location where a 3D null point could appear. Potential null point structures are very resilient to numerical resolution \citep{2017A&A...606A..46G}, while the increase in free magnetic energy could collapse the null points into current sheets that may make it much more difficult to identify \citep{1997JGR...102..231G}. In the BP10 case a 3D null point is identified that is likely to be the key for the magnetic energy release process associated with this eruption.

A recent investigation by \citet{2015JApA...36..123P} shows that the paths of erupting prominences (large CMEs) are to a large degree determined by the structure of the overlying coronal magnetic field. The CMEs in their investigation are seen to follow the path through the magnetic field that offers the least `resistance'. This implies that the CMEs are most likely to propagate along the weak open field (OF) line regions. With this knowledge, the propagation of the mini-CME seen in the imaging observations provides indirect information on the structure and possible field strength of the unseen coronal magnetic field. The structure of the ejecta seen in the observations may in some cases be explained by investigating the structure of the large scale 3D magnetic field from the NLFFF relaxation simulation. Here, the ejecta is either propagating along the existing magnetic field lines, or expanding along arcade structures.  The eruptions or mini-CMEs discussed above are small scale versions of the large and more impressive CMEs. As in the case of classic CMEs, the eruptive structures are most likely to expand along a channel of weak magnetic field strength.

 The advantage of studying mini-CMEs relates to the simplicity of the underlying magnetic field. In the CBP mini-CME cases the photospheric magnetic field structure is simple, with only a relatively small number of strong magnetic flux concentrations defining the local magnetic field structure. In comparison, large CMEs are typically associated with active regions that are highly complex magnetic systems. It is therefore much easier to understand the magnetic structure and the reasons for both the formation and eruption of mini-CMEs. This makes it very valuable to continue investigating these events, and to increase the spatial resolution of future observations to better capture the magnetic flux and emission distribution. Knowledge of the basic processes in the mini-CMEs may subsequently be used to improve our understanding of the evolution of large CMEs.

 All of the events in our sample refer to non-emergence cases. In Paper~I we found that  ``from 21 eruptions, 18 occurred 9 to 25~h after the CBP formation, with an average delay of 16~h 40~min. This time delay in the eruption occurrence coincides with the CBPs become smaller until they fully disappear.'' In the remaining three cases the first eruptions occurred relatively soon after the CBP formation ($\sim$5~h, BP1, 4~h, BP7, and 6~h, BP8), i.e. the CBP appearance in the AIA 193~\AA\ channel.  A recent paper by \citet{2018ApJ...852...98W} is interesting in this context. An initial open magnetic field contains an opposite polarity flux region, with a single 3D magnetic null point above the photosphere, where the null points spine axis connects to the minority flux polarity below it. The single polarity magnetic field is stressed by a twisting motion that follows the contours of the Bz component of the minority polarity region. This builds up stress, forming a flux rope that eventually erupts, producing the ejection of twisted plasma to the outside open flux region. A process the authors refer to as a breakout model. The above model is a modification of the model discussed in \citet{2009ApJ...691...61P, 2010ApJ...714.1762P}, where the difference is the formation of a twisted filament inside the null point dome region that becomes unstable. The problem in these studies relative to the observations is the huge amount of twist needed to reach the unstable state. This twisting seems not to be present in any of the observed cases investigated here.

In many observations the eruptions have been inferred to as a consequence of the general cancellation of magnetic flux in the photosphere. One may make time dependent measurements of the signed magnetic flux in the photosphere over long time periods, and these often show how the signed flux in a specific domain decreases systematically over long time scales. The fact that the change in flux takes place over timescales much longer than the eruption event, renders this interpretation of the trend misguiding and it must be due to other processes of a much less explosive character. 

The observations considered here show that all eruptions are initiated from a confined area in the magnetic structure, which is most likely located below the twisted flux concentrations. From a modeling perspective there are different possibilities. Either a traditional kink instability that requires the magnetic field strength to decrease significantly outside the flux rope. For the cases considered here this is not likely as in the model magnetic field, the gradients in magnetic field strength are not large. A second possibility is a buoyancy instability that lifts the twisted loop upwards  from low down in the atmosphere, creating a vertical current sheet below the  flux rope in the lower corona. Tether cutting can then drive the process explosively, as it has been seen in previous cases \citep{2004ApJ...610..588M, 2013ApJ...771...20M}, where it has an important role in both propelling the overlying  flux rope upward and adding new twist to the underlying magnetic field, helping build up twist for a following eruption.

\section{Conclusions}
\label{sect:sum}

The time dependent evolution of a series of eruptive CBPs have been investigated using a NLFFF simulation approach. An initial potential solution to the observed HMI magnetogram is advanced in time using subsequent HMI magnetograms with a time resolution of 450 seconds. The time dependent NLFFF solution is found using a relaxation approach, which allows complicated field line structures to build up over time. The magnetic field configurations have been investigated around the times where the CBPs erupt, identifying the magnetic field configuration responsible for the eruption and investigating the field for the possible paths of the ejecta. In 17 out of 21 cases a twisted magnetic field structure has been located in the close vicinity of the first intensity enhancement in the AIA observations. This indicates that the existence of magnetic flux ropes may be the main ingredients in the eruptions, and that these events may be comparable in structure and initial evolution to large CMEs. 

The structure of the large scale magnetic field differs significantly between the different CBP studies. In two cases a FS like structure without any identified null points are identified. Seven CBP cases have LASs. Two cases have an OF structure and finally two contain a more complicated field line structure. In all cases the observed eruptions take place on the edge of the central CBP loop system, where one of the CBP flux concentrations connects in a SsAS structure to a nearby opposite polarity flux concentration. 

From Paper~I it is seen that repeated eruptions take place in a large fraction of the selected CBPs, and the modeling here shows that these occur in the same topological region of the magnetic field and therefore can be classified as homologous eruptions. These signatures of the CBP eruptions need to be explained. This includes an understanding of why these regions are susceptible to the build up of magnetic twist and free energy, which eventually becomes unstable, leading to the observed eruptions. To reach this goal two advances are required. At present the CBP regions are only marginally resolved in the HMI observations. This relates both to the spatial resolution and to the limiting threshold of the magnetograms. The lack of resolution influences the results from the numerical modeling. At present the small scale magnetic features that are important for the local twisting of the magnetic field are represented by only a few pixels. This implies that the numerical model operates on a length scale relative to the magnetic structures that is close to the diffusion length scale. This has a negative influence on the results in the form of high numerical diffusion in these regions, which limits a realistic evolution of these regions. To follow and understand the ongoing field evolution more realistically, higher space and sensitivity resolution data and preferably vector magnetograms are needed.

The question about the true nature of the micro-flares (compact sudden brightening associated with PIL that follow/accompany the start of the eruptions) remains unclear. A follow-up study on a coronal-hole CBP eruption case may provide a conclusive answer on this issue. 

\section*{Acknowledgments}
 We thank the referee for the suggestions to improve the paper.
VAPOR is a product of the National Center for Atmospheric Research's Computational and Information Systems Lab. Support for VAPOR is provided by the U.S. National Science Foundation (grants $\#$ 03-25934 and 09-06379, ACI-14-40412), and by the Korea Institute of Science and Technology Information. C. M. thanks the National Natural Science Foundation of China (41474150).
The HMI data are provided courtesy of NASA/SDO and corresponding science teams. The HMI data have been retrieved using the Stanford University's Joint Science Operations Centre/Science Data Processing Facility. M.M. and K.G. thank the ISSI Bern for the support to the team ``Observation-Driven Modelling of Solar Phenomena''.

\bibliographystyle{aa}
\bibliography{refs}

\begin{appendix}

\section{Details of the remaining BPs}
\label{details}

Here the individual eruptions that are not included in the main text are discussed in detail. Modelling movies for the left panels of the figures shown below are found here \url{https://doi.org/10.5281/zenodo.1915477}. The movies are named as BP?\_erupt?.mov, where the first ? represents the CBP case number and the second ? the eruption number.

\begin{figure*}[h]
{\hfill
\includegraphics[scale=0.23]{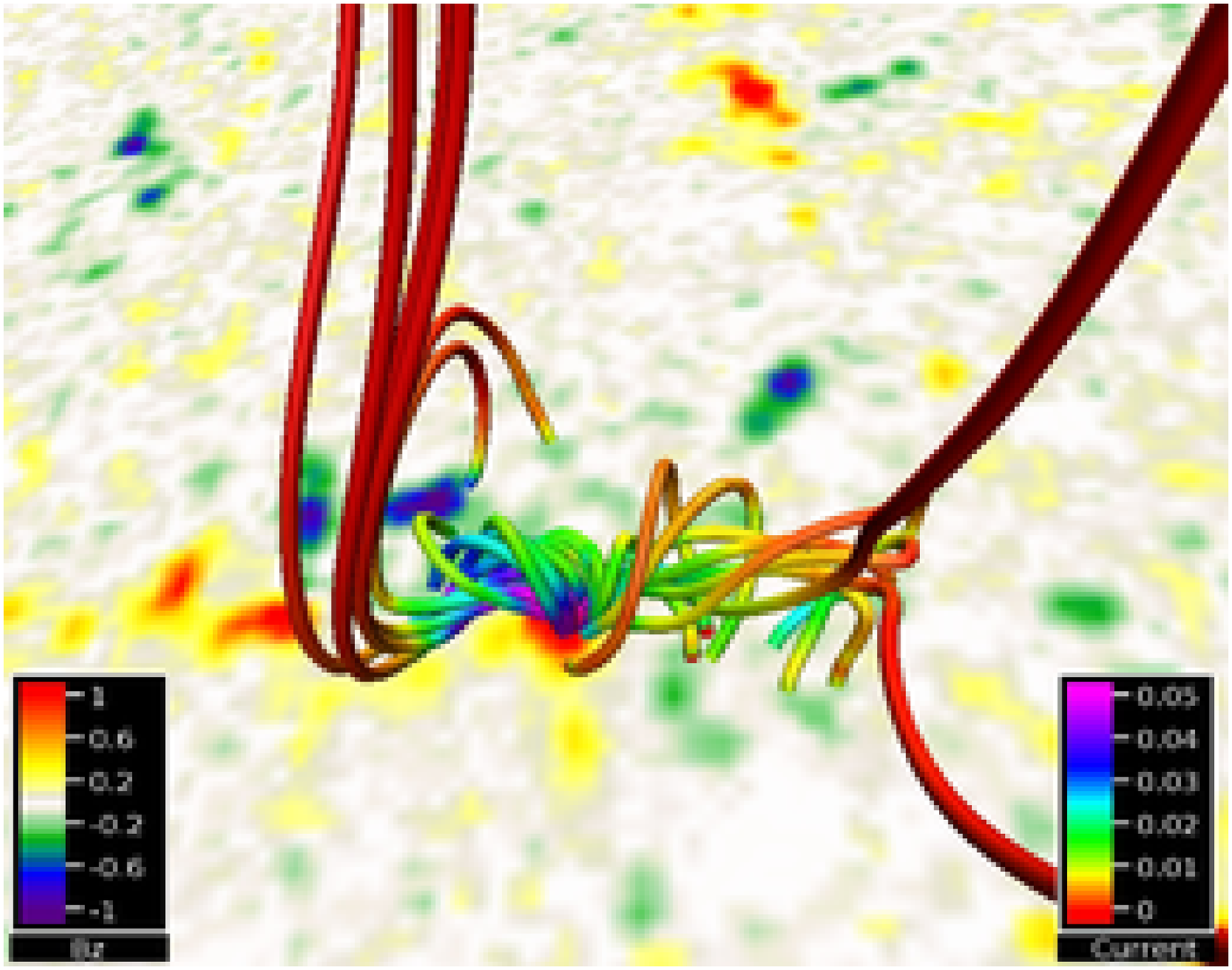}
\hfill
\includegraphics[scale=0.23]{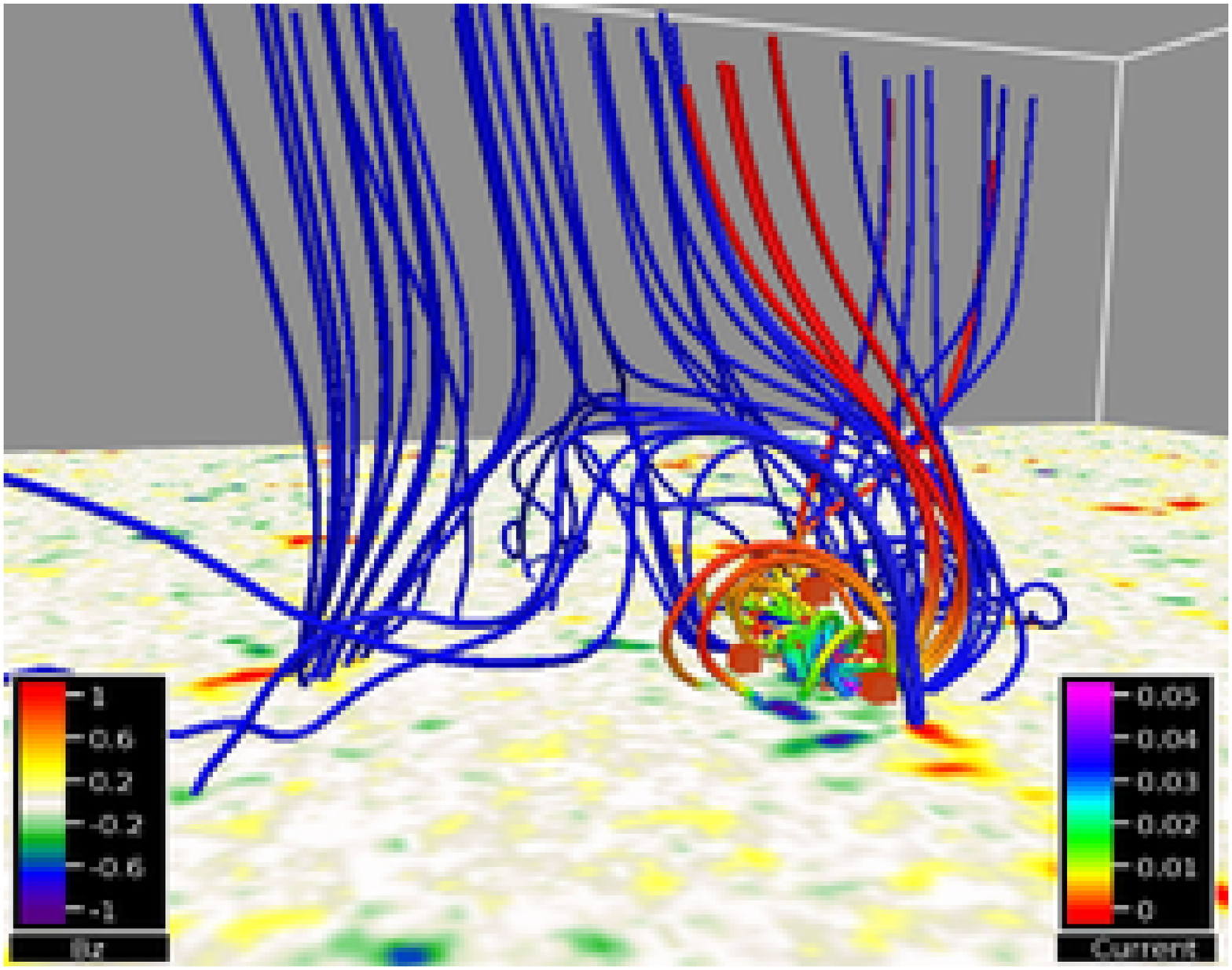}
\hfill}
\caption[]{As in Fig.~\ref{BP1_1} but for BP2--ER1. Left panel shows the combined sheared and twisted magnetic field line structure that may constitute the unstable field that is triggering the eruption. The right panel shows with blue field lines the large scale open magnetic field with the twisted structure embedded in it. Time: 19:30 UT.}
\label{BP2_1}
\end{figure*} 


{\it BP2--ER1:} This eruption takes place in the close vicinity of two very closely placed opposite polarity flux concentrations at the west edge of the CBP region. A detailed investigation of the field line structure of this region shows two interesting features. First, the local field lines generally contain lots of current and secondly, these field lines form two  FRs that extend in both the east and west directions relative to the local dominating strong positive flux concentration. Further to this, there is a SsAS connecting the two dominant positive and negative flux concentrations, that helps to keep the twisted field lines confined in height, left panel of Fig.~\ref{BP2_1} (See the movie associated with Fig.~\ref{BP2_1}). From the time evolution it is seen that the field lines maintain their high current values throughout the time series, giving little clue to when a possible eruption may take place. The magnetograms show various motions of the flux concentrations over the time scale of the observations, with a trend to make the flux distribution simpler, resulting in fewer and stronger flux concentrations scattered over a smaller region. 
Upon considering the large scale magnetic field structure, one sees that the MF is located in one of two magnetic field loop systems contained inside a  FS like structure, similar to the BP1 case (see the right panel of Fig.~\ref{BP2_1}). The  top `spine' FS connects towards the top of the simulated domain due to a general flux imbalance of the applied HMI magnetogram region.
The magnetic field strength decreases with height, where it is found to be marginally stronger to the south of the twisted structure. From the simulated magnetic field, it is surprising that the eruption expands in the southern direction. It seems to be energetically easier for the eruption to move upwards and follow a slightly northern direction by expanding up through the OF. 

\begin{figure*}[]
{\hfill
\includegraphics[scale=0.23]{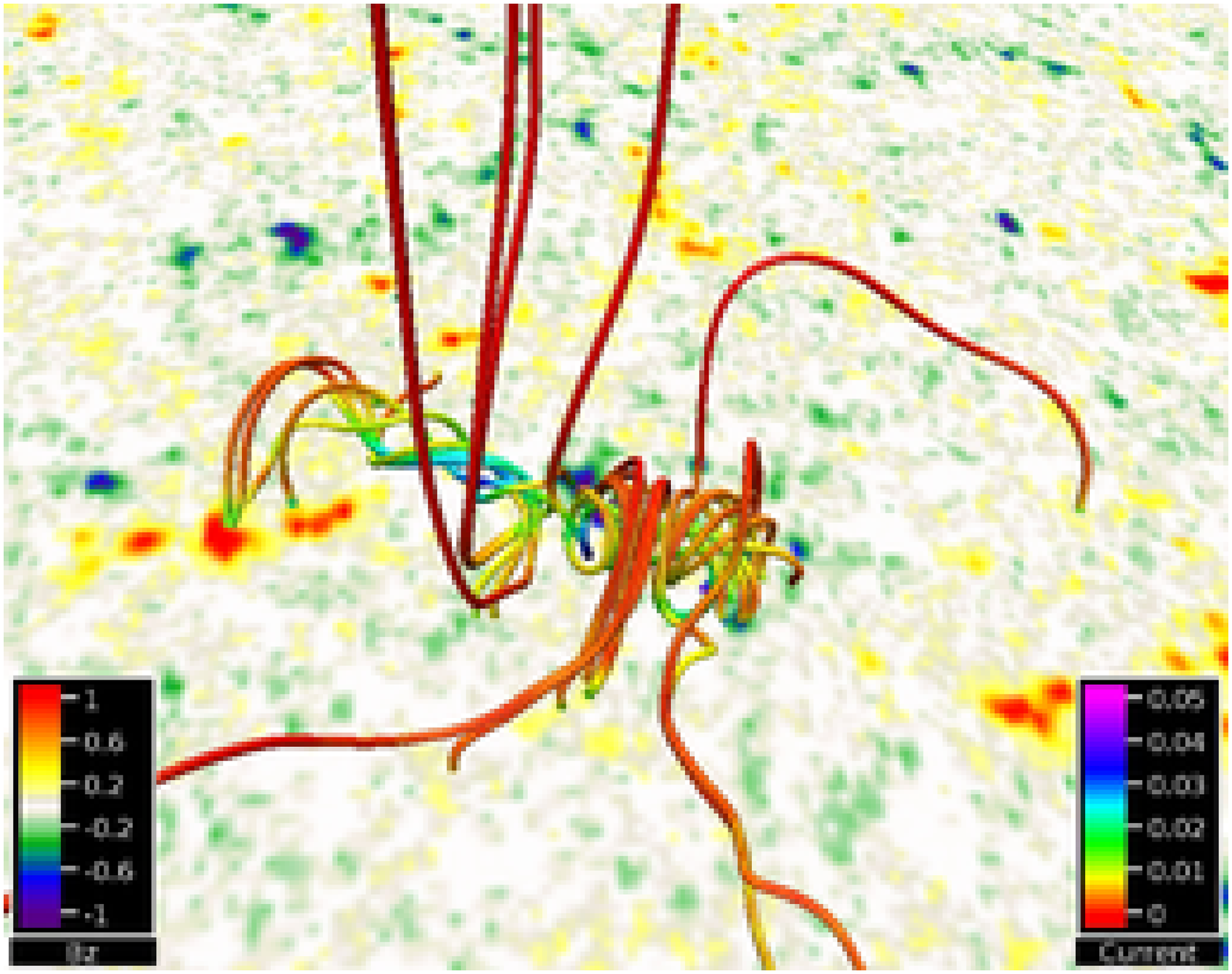}
\hfill
\includegraphics[scale=0.23]{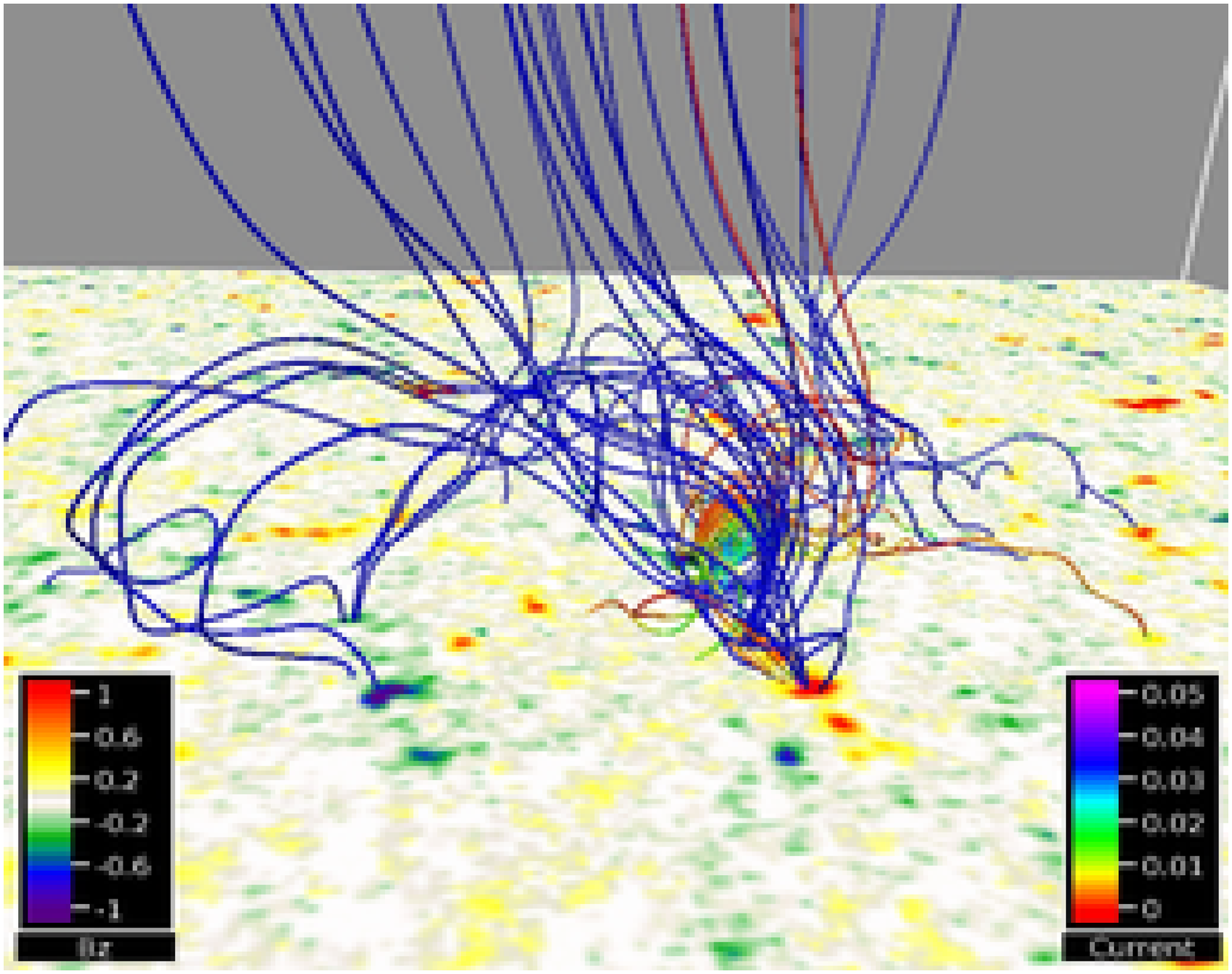}
\hfill}
\caption[]{As in Fig.~\ref{BP1_1} but for BP2--ER2. Time: 23:09 UT.}
\label{BP2_2}
\end{figure*}

{\it BP2--ER2:} The simulation data show a  FR associated with the two closest opposite flux concentrations at the source location of the eruption with a preference for the FR to extend in the east direction, left panel of Fig.~\ref{BP2_2} (see the movie associated with Fig.~\ref{BP2_2}). At the west end of the FR, a separator surface limits its extension. Additionally, a twisted structure extends to the north-east from the local twisted flux region and connects to the nearby positive flux concentrations. The FR contains a relatively high amount of current, indicating available free magnetic energy. Due to the weaker negative flux imbalance over the global area of interest, a large weak FS like structure embeds the twisted field regions, right panel of Fig.~\ref{BP2_2}. Seemingly the centrally placed twisted magnetic field causes the eruption, which takes place in the remnants of the magnetic field environment of the first eruption, making this a homologous event. The ambient magnetic field is open above a given height that would permit some of the erupting material to escape.


\begin{figure*}[]
{\hfill
\includegraphics[scale=0.23]{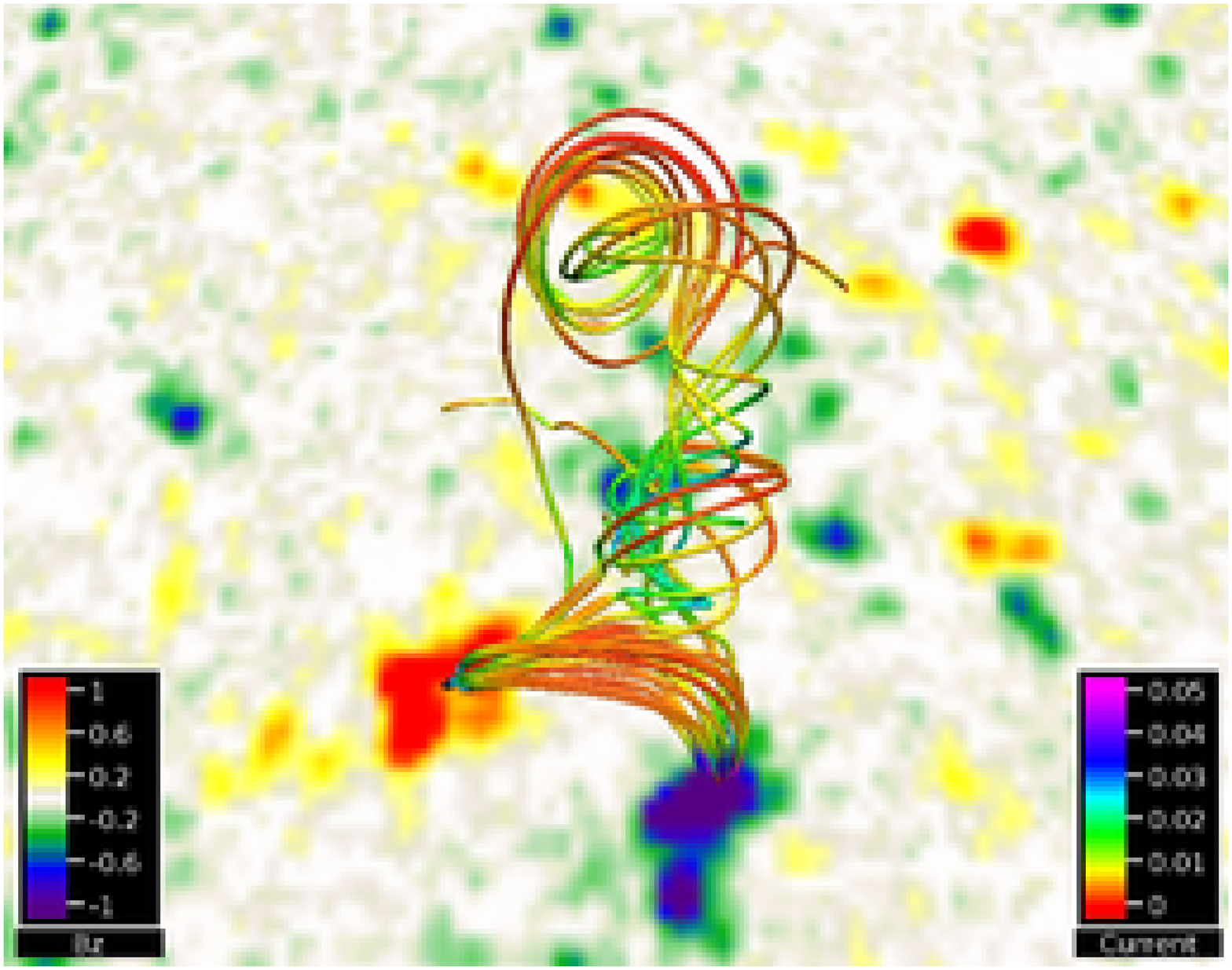}
\hfill
\includegraphics[scale=0.23]{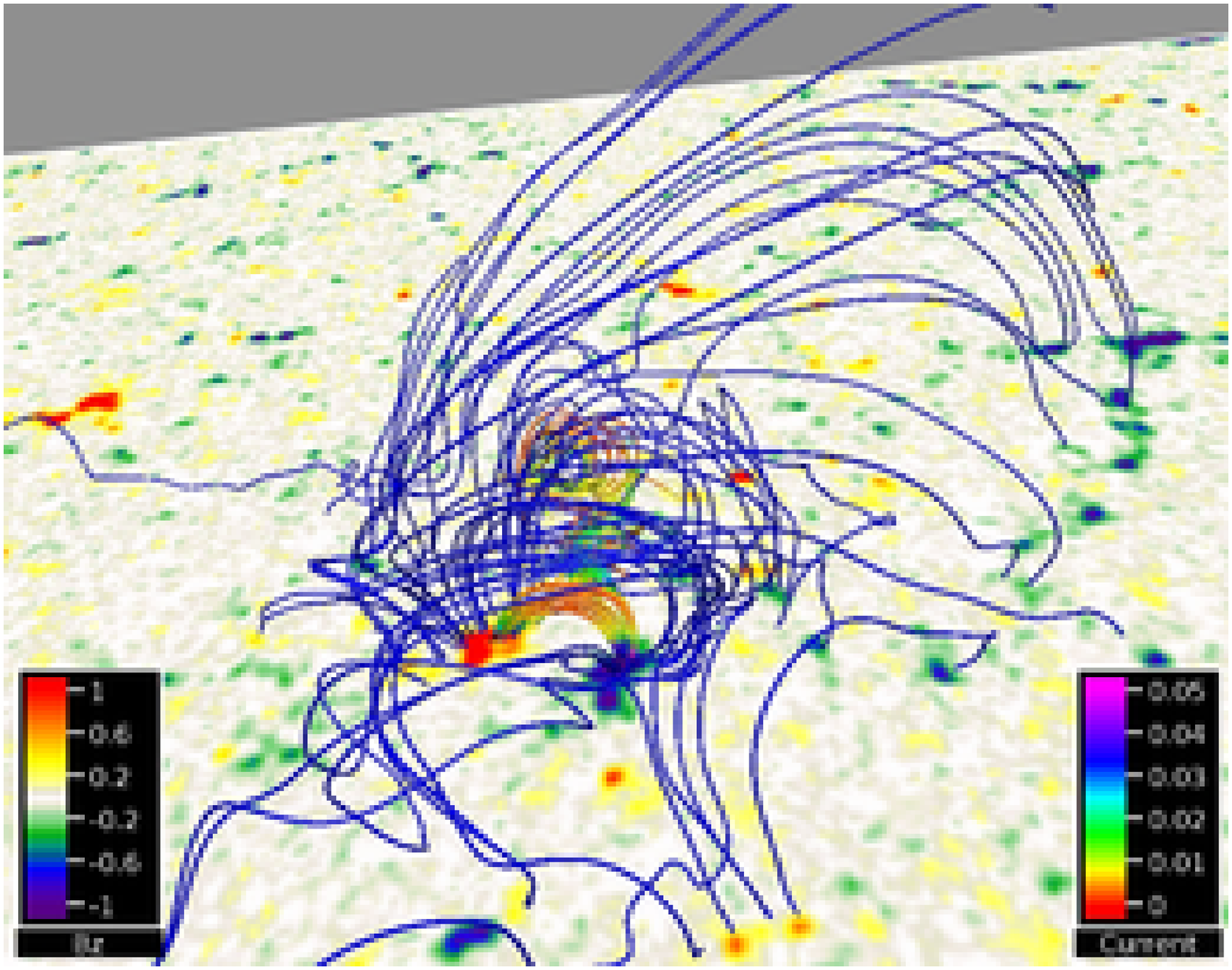}
\hfill}
\caption[]{As in Fig.~\ref{BP1_1} but for BP3--ER1. The left panel shows the arcade structure connecting the two main polarities of the core CBP region and twisted field line structure to the north of this that may constitute the unstable field that triggers the eruption.  Time: 09:21 UT.}
\label{BP3_1} 
\end{figure*}

{\it BP3--ER1:} The magnetic field simulation from the NLFFF relaxation simulation shows a SsAS connecting the two main flux concentrations in the CBP region and a FR connecting to the north of the main CBP region passing above the negative flux concentrations (left panel of Fig.~\ref{BP3_1} and the movie associated with Fig.~\ref{BP3_1}). This northward connection is clearly seen in all AIA channels and it is where the eruption takes place. The  FR continues even further to the north connecting to a smaller cluster of positive flux concentrations. This extension is not observed in the AIA images, where the enhanced emission seems to end at the first negative flux concentration. The twisted magnetic field has only a weakly enhanced current compared to the previous cases and is found to change structure and concentration with time over the period of the eruption. The  FR is embedded in a LAS that is orientated in the east-west direction. From the field line trace in the right panel of Fig.~\ref{BP3_1}, it is clear that the lower field lines confining the  FR are rooted in the photosphere on either side of the  FR. Field lines passing higher above the FR connect from the dominating positive flux to a variety of negative flux concentrations on the order of 30\arcsec\ to the west of the FR. This indicates that the energy release in this case is contained within the closed magnetic structure identified in the model data. 


\begin{figure*}[]
{\hfill
\includegraphics[scale=0.23]{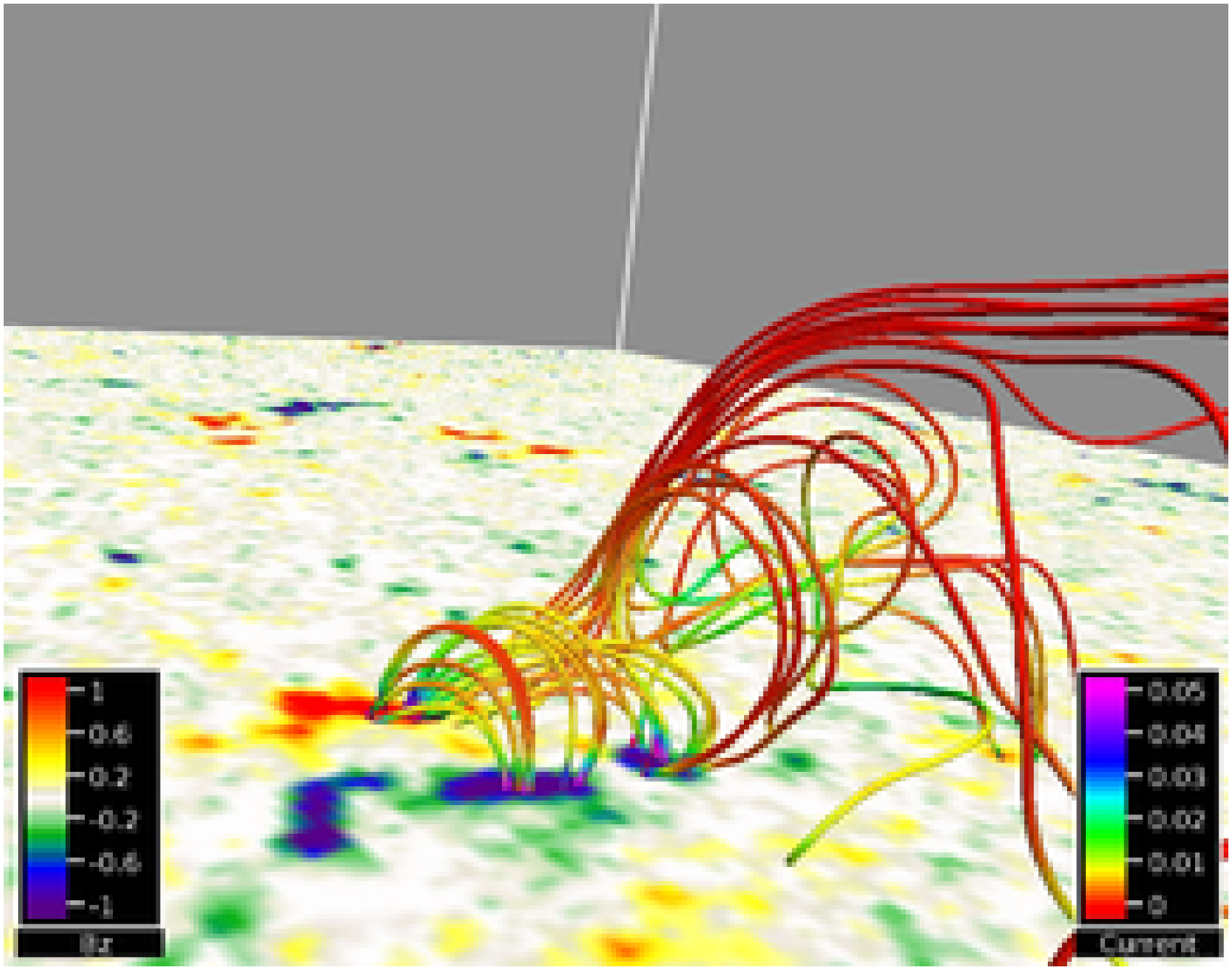}
\hfill
\includegraphics[scale=0.23]{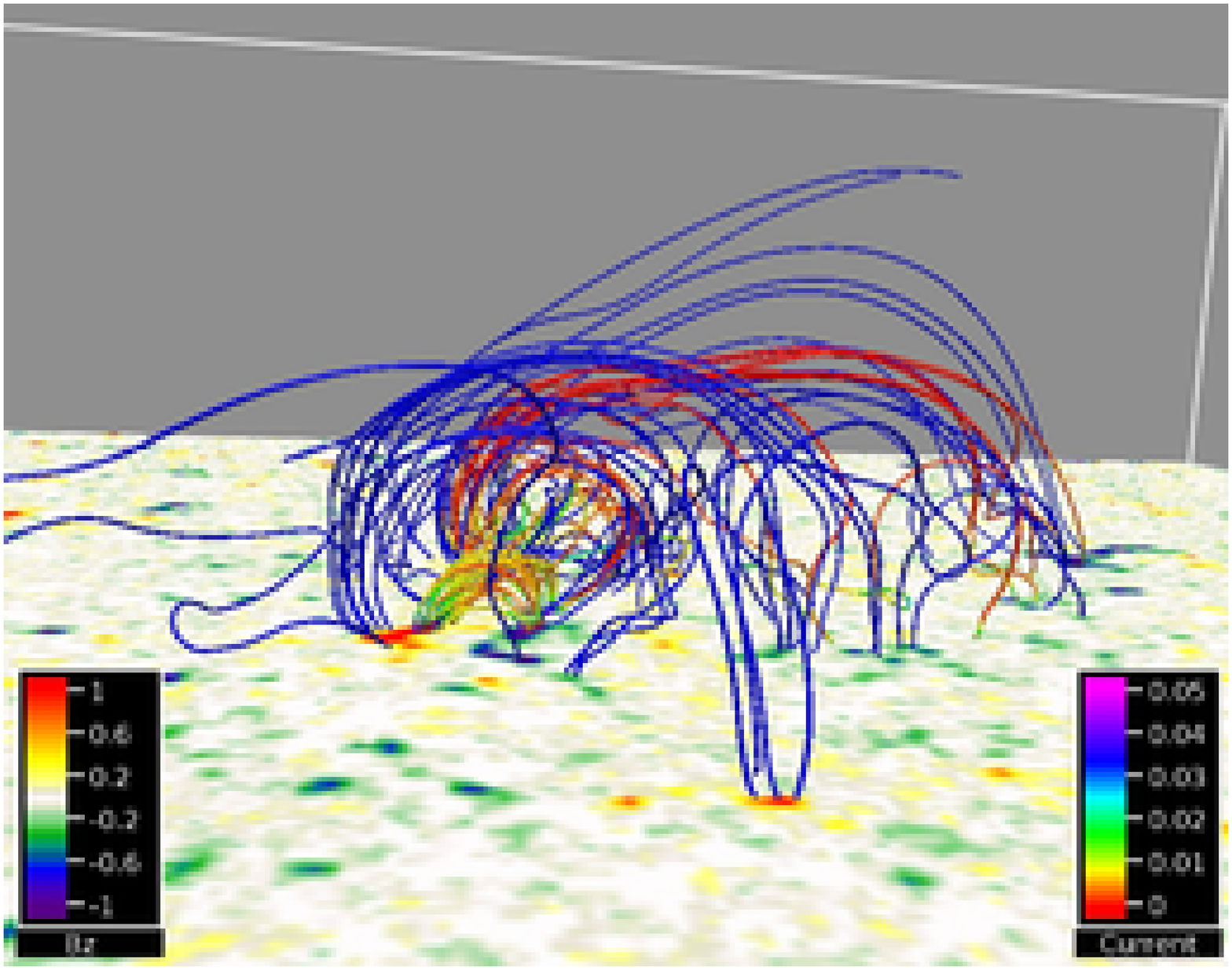}
\hfill}
\caption[]{As in Fig.~\ref{BP1_1} but for BP3--ER2.  Time: 13:31 UT.}
\label{BP3_2} 
\end{figure*}

{\it BP3--ER2:} The second eruption of BP3 takes place in the same magnetic configuration as the first eruption. A difference from the previous eruption, that is 4 hours earlier, is that this eruption is ejected in the north-east direction. The magnetic structure of the underlying photospheric field has evolved from that seen earlier and now consists of a row of positive and negative fluxes with a slight sheared structure between them. The CBP region is mainly associated with the positive polarity and the more southern negative flux concentrations, with a clear connection between the northern part of the positive flux and the most northern negative flux concentration (see Fig.~\ref{BP3_2} and the movie associated with Fig.~\ref{BP3_2}). The left panel of Fig.~\ref{BP3_2} shows the structure of the magnetic field, with a clear sheared arcade covering a large fraction of the extended bipolar region. Over time the distance between the rows of positive and negative flux concentrations decreases, while there is also an internal reconfiguration of their relative positions and flux concentrations. This process continues after the start of the eruption. No strongly twisted or sheared field lines are identified inside the arcade which hosts the eruption up to the time of the energy release. To the north of the main CBP region there are loops reaching 3--4 times higher into the atmosphere, see the right panel of Fig.~\ref{BP3_2}. Comparing this figure with the same for the BP3--ER1 event, it can be seen that the large scale structures are comparable between the two events. 


\begin{figure*}
{\hfill
\includegraphics[scale=0.23]{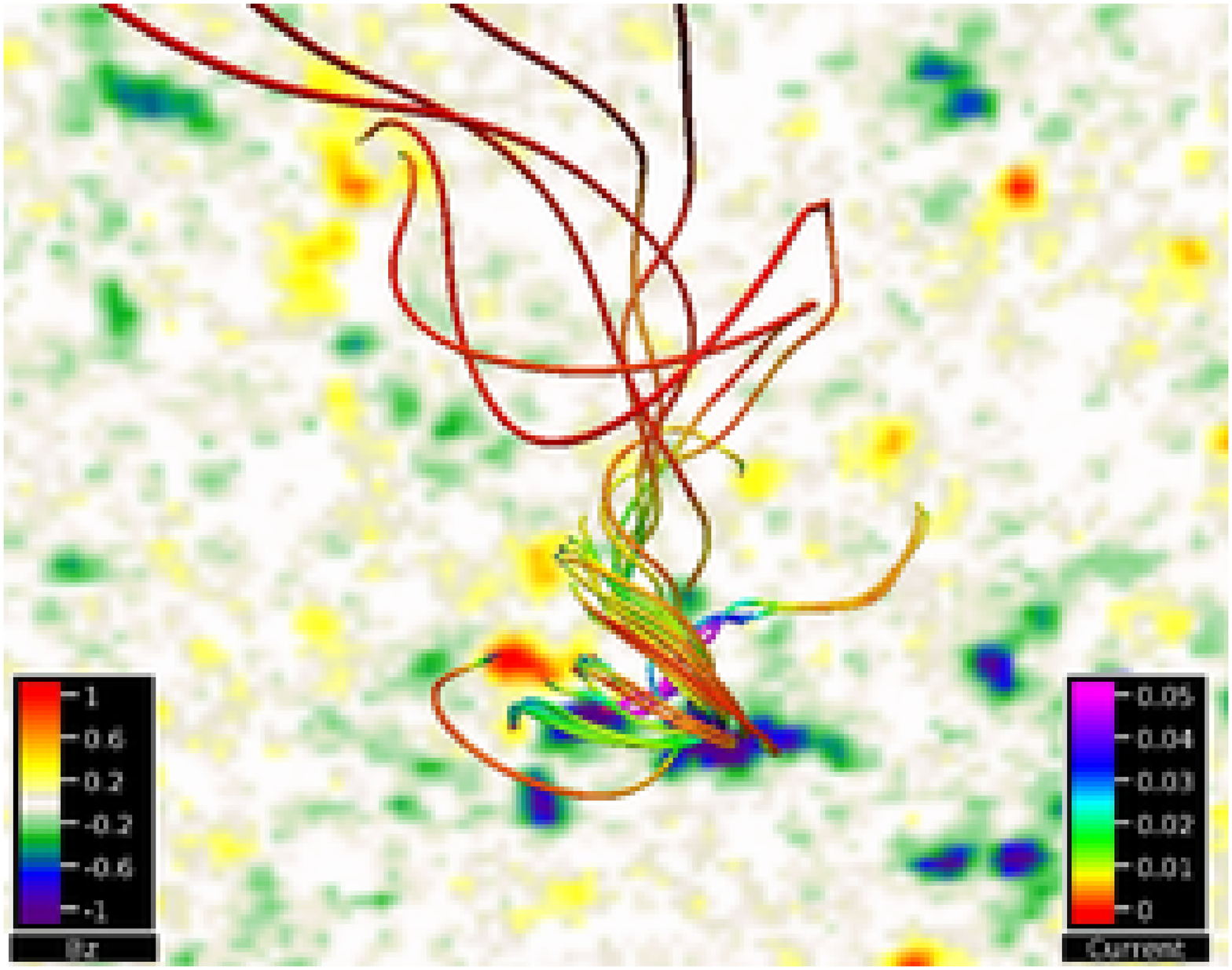}
\hfill
\includegraphics[scale=0.23]{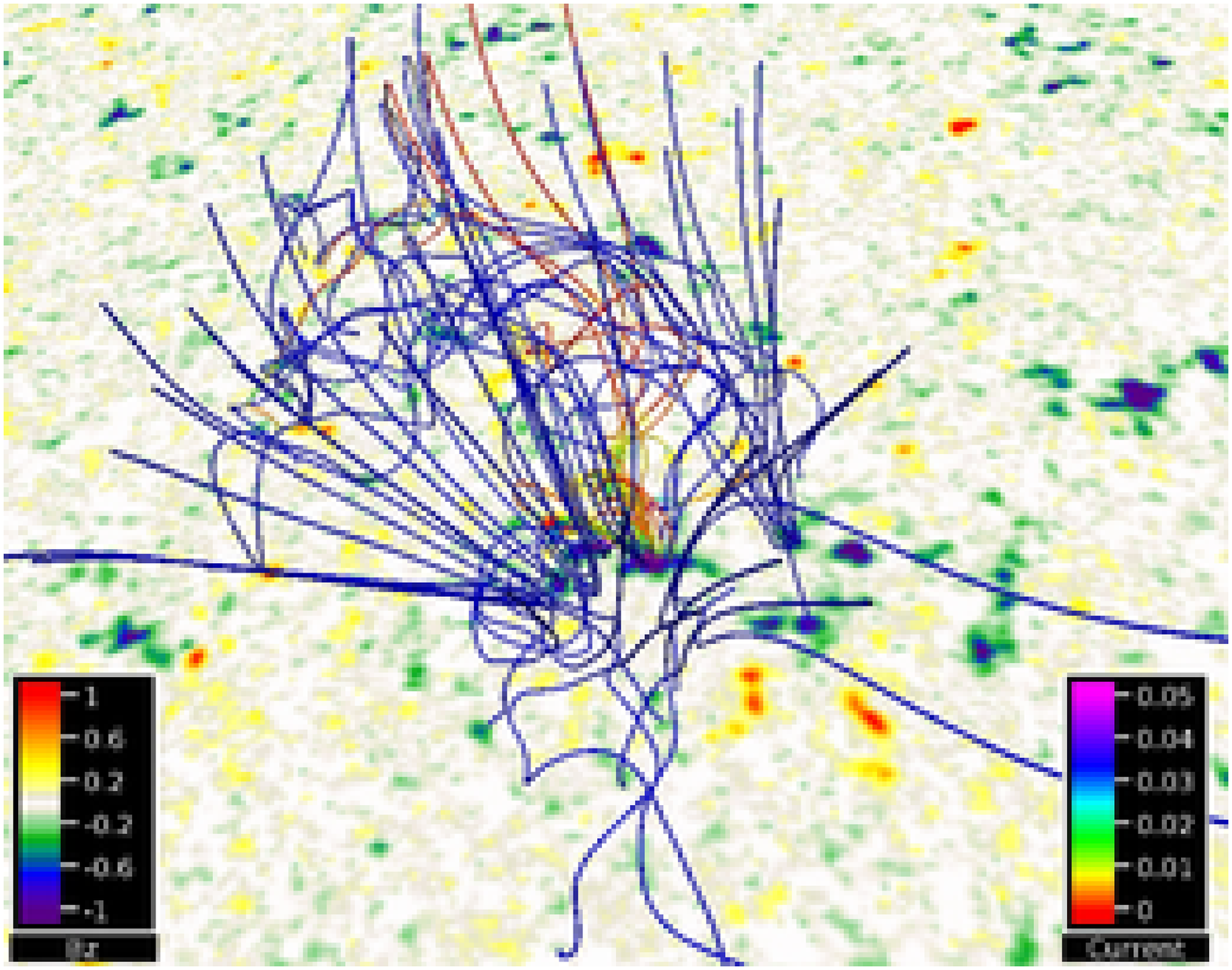}
\hfill}
\caption[]{As in Fig.~\ref{BP1_1} but for BP4--ER1. Time: 03:25 UT.}
\label{BP4_1}
\end{figure*}

\begin{figure*}
{\hfill
\includegraphics[scale=0.23]{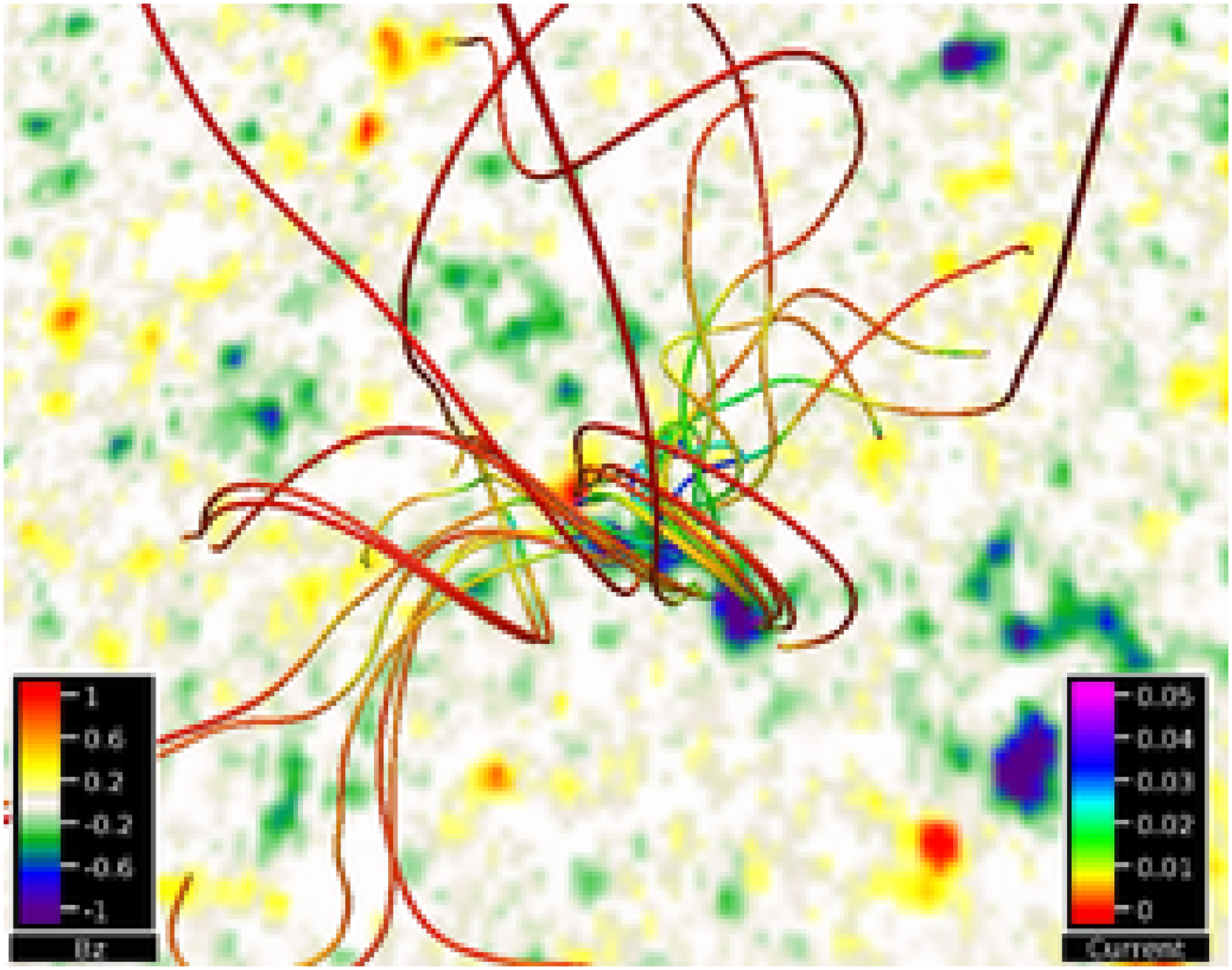}
\hfill
\includegraphics[scale=0.23]{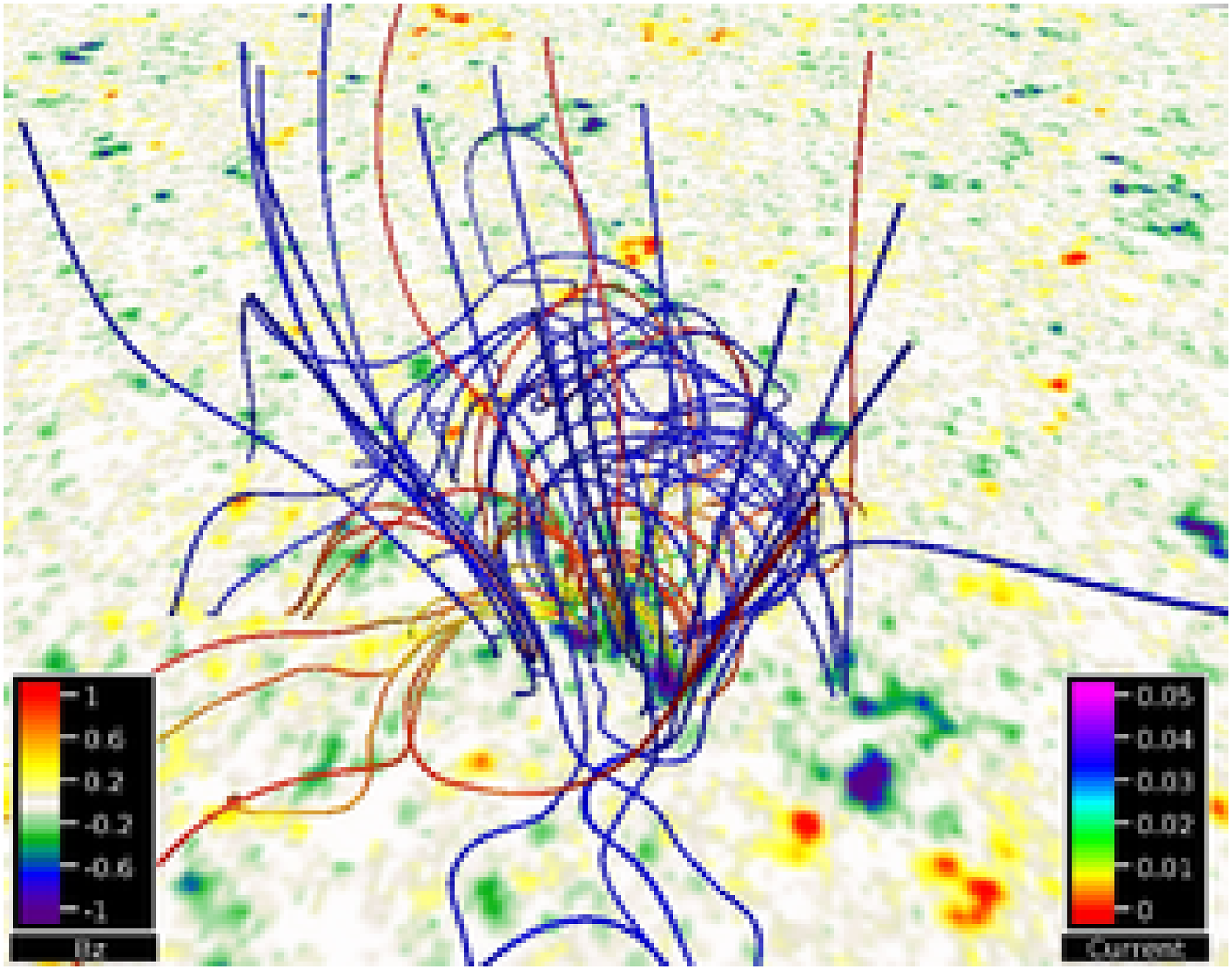}
\hfill}
\caption[]{As in Fig.~\ref{BP1_1} but for BP4--ER2.  Time: 05:15 UT.}
\label{BP4_2}
\end{figure*}

{\it BP4--ER1:} The magnetic field structure represents a SsAS connecting the two main opposite flux concentrations. South-west of these is a second negative flux concentration. Connections between these three flux concentrations generate a more complicated SsAS system. From the most eastern negative flux concentration, magnetic flux also connects towards the north, generating a  FR that is located close to the point where the eruption initiates, left panel of Fig.~\ref{BP4_1}. The different field line connectivity regions impose a location with a high shear in the field, that may allow for a small energy release (see the movie associated with Fig.~\ref{BP4_1}). The large scale structure of the field, right panel of Fig.~\ref{BP4_1}, represents several LASs connecting between the different positive flux concentrations to more diffuse negative flux locations. This defines a loop system extending in the north-south direction, with a slight turn in the east direction as the distance north of the CBP region increases.


{\it BP4--ER2:} The eruption takes place in the same region of the magnetogram as the previous eruption, although some modifications of the photospheric magnetic field have taken place. The change is related to the re-orginasation of the flux concentrations to new relative positions and a continued simplification of the surface flux distribution. This event is another example of a homologous eruption. There is a  FR connecting in the north direction from the negative flux concentrations, left panel of Fig.~\ref{BP4_2} (see the movie associated with Fig.~\ref{BP4_2}). This  FR is close to the position where the initial emissivity is seen in the  AIA 171 and 193~\AA\ observations. The twisted structure contains a medium strong current located in the core region of the eruption. The large scale magnetic field has a close resemblance to the field seen in BP4--ER1 (a LAS), right panel of Fig.~\ref{BP4_2}. 
 

\begin{figure*}
{\hfill
\includegraphics[scale=0.23]{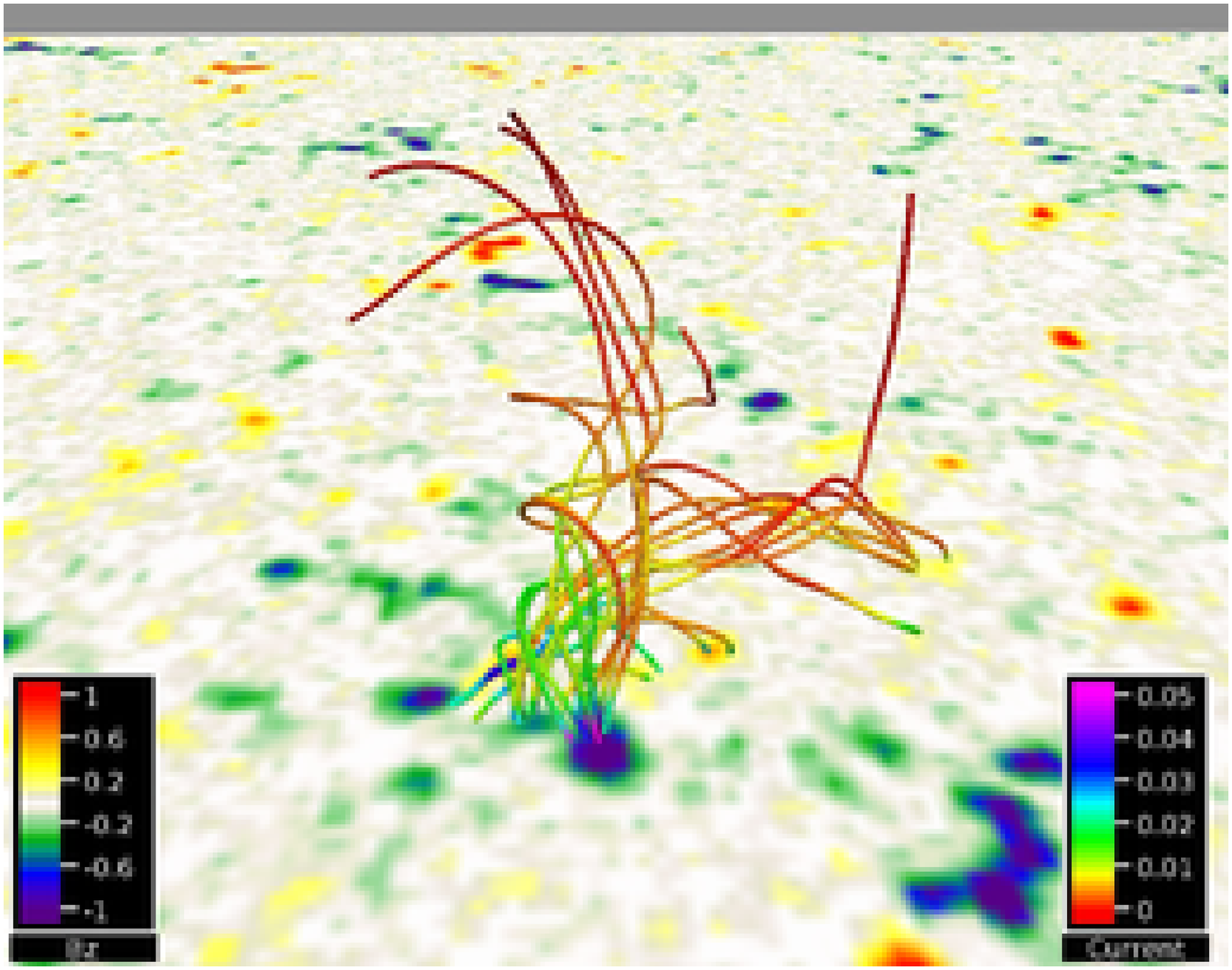}
\hfill
\includegraphics[scale=0.23]{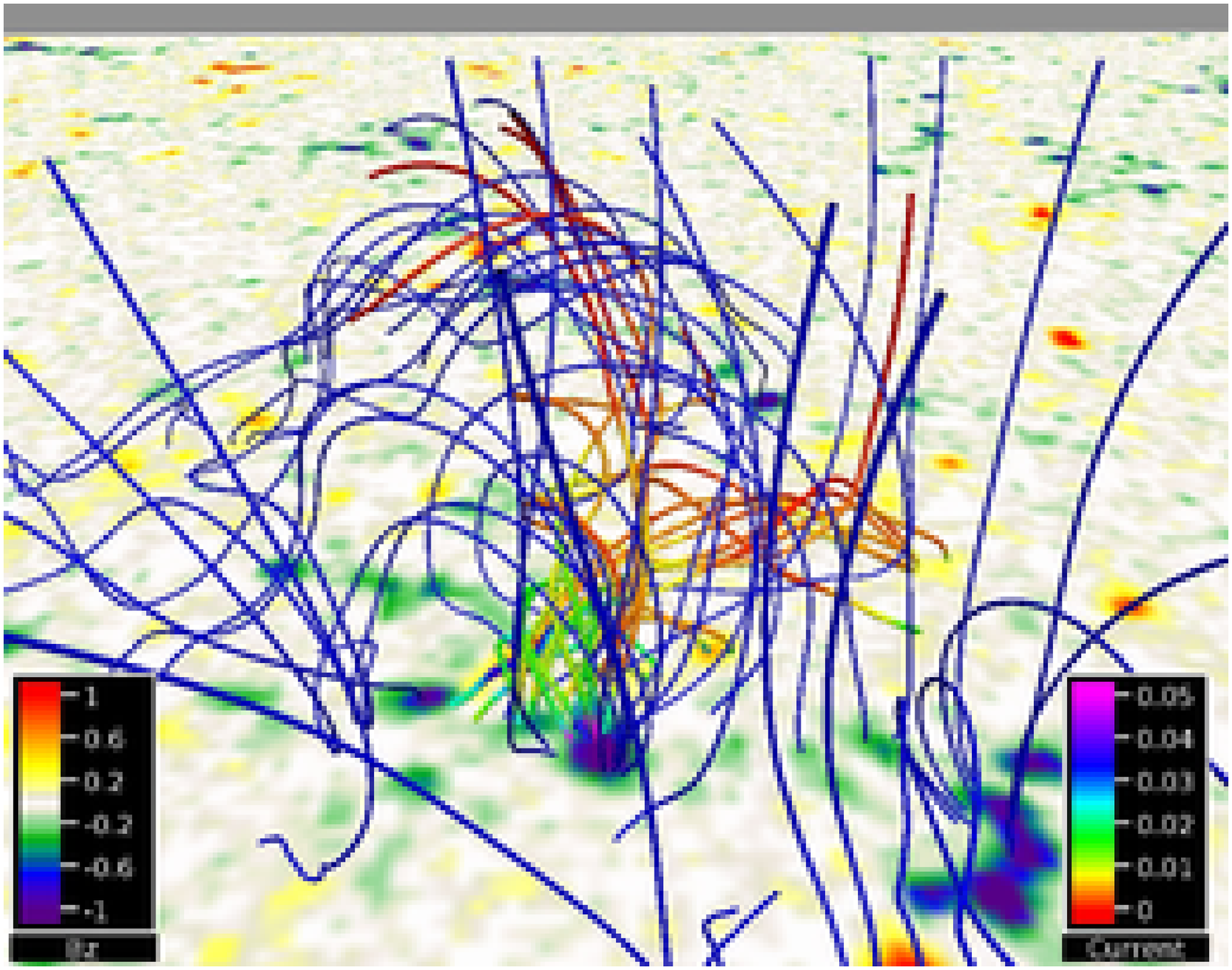}
\hfill}
\caption[]{As in Fig.~\ref{BP1_1} but for BP4--ER3. Time: 06:15 UT.}
\label{BP4_3} 
\end{figure*}

{\it BP4--ER3:} The location of the eruption is similar to the two previous eruptions, and the structure of the ejection resembles the two previous cases. The NLFFF simulation data again show a  FR where one footpoint starts from the negative flux concentrations extending northward. This time the structure is smaller and more difficult to identify, left panel of Fig.~\ref{BP4_3} (see the movie associated with Fig.~\ref{BP4_3}). Over time the direction of the twisted flux concentration slowly changes from a near north direction to a north-west direction. A partial reason for this relates to the re-organisation in the flux concentrations, with the minority positive flux moving towards the west while slowly vanishing over time. The LAS represents the same general structure that was seen for the two previous cases, right panel of Fig.~\ref{BP4_3}. 

\begin{figure*}
{\hfill
\includegraphics[scale=0.23]{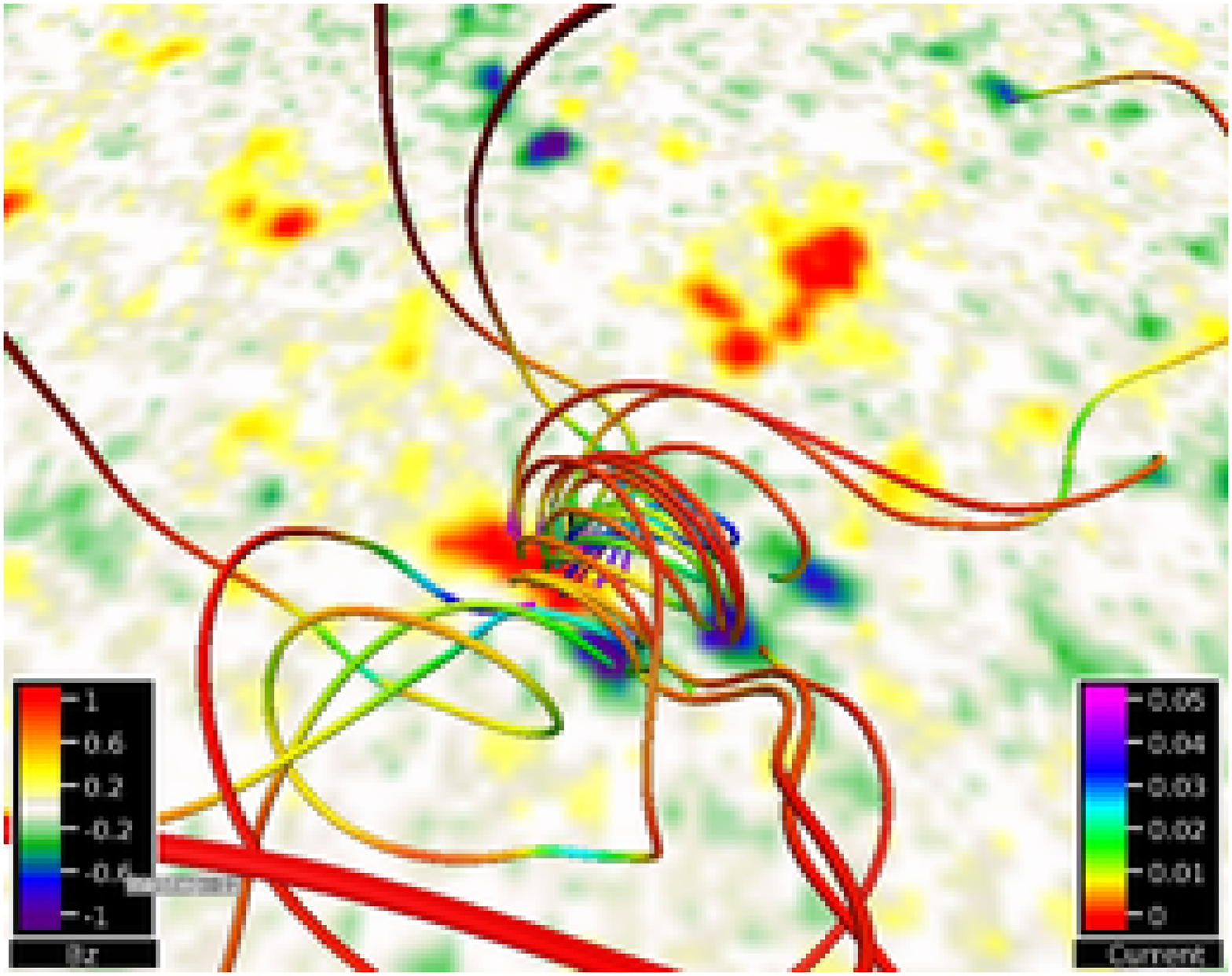}
\hfill
\includegraphics[scale=0.23]{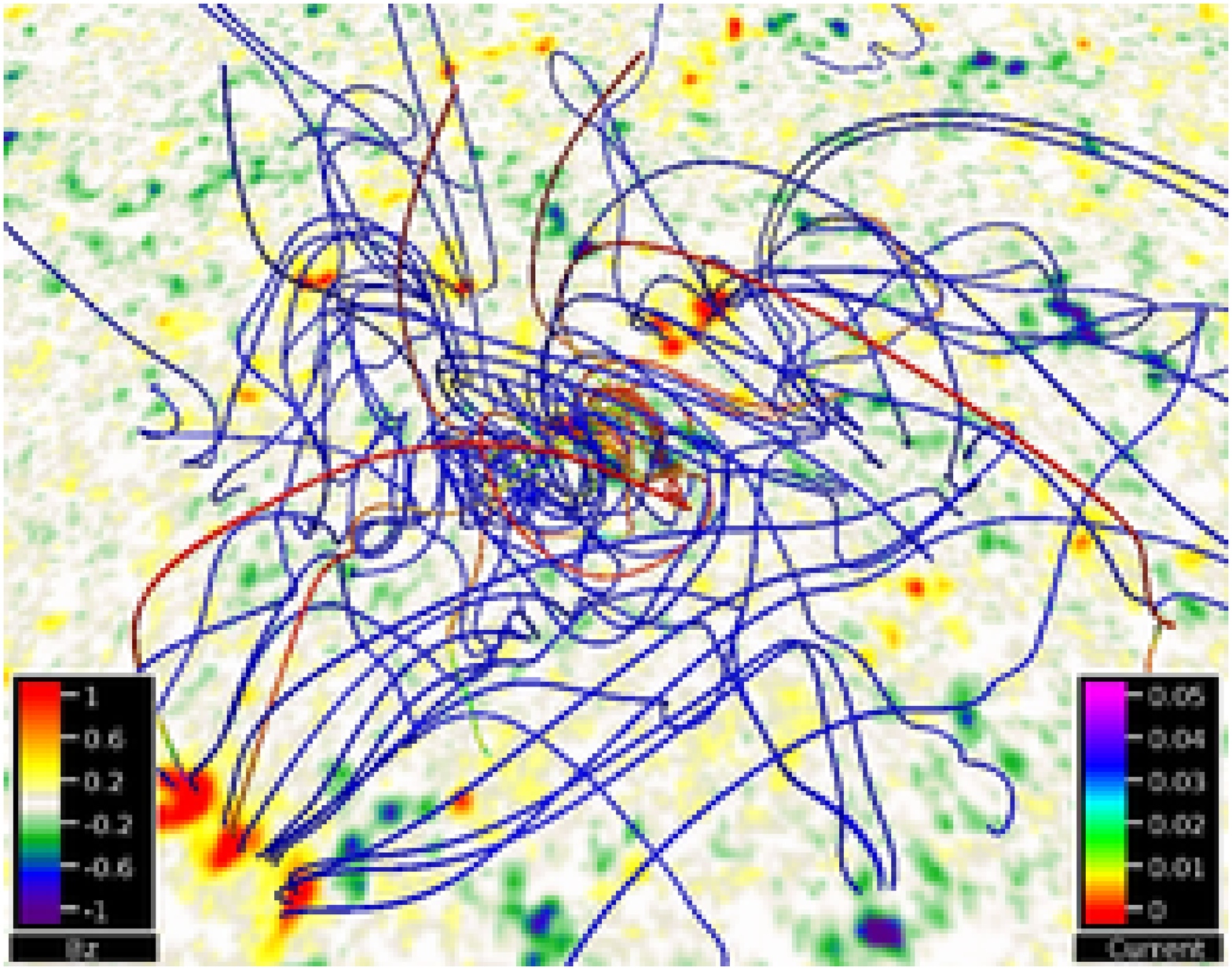}
\hfill}
\caption[]{As in Fig.~\ref{BP1_1} but for BP5. Time: 21:34 UT.}
\label{BP5}
\end{figure*}

{\it BP5:} The eruption takes place from a small region where a single large positive flux concentration and several smaller negative concentrations are present. The local field line structure shows a SsAS connecting between the positive and negative flux polarities defining the core region of the CBP. To the south-east of these flux concentrations one finds a  FR that extends to a nearby larger positive flux concentration, left panel of Fig.~\ref{BP5} (see  the movie associated with Fig.~\ref{BP5}). The  FR contains relative strong current over a larger region making it a candidate for an eruption. The  FR extends along the direction of the eruption when compared to the AIA observations. The large scale magnetic field consists of a LAS that confines the  FR, and connects a significant fraction of the field lines in the south-east direction towards a few distant positive flux concentrations, right panel of Fig.~\ref{BP5}. 

\begin{figure*}[!h]
{\hfill
\includegraphics[scale=0.23]{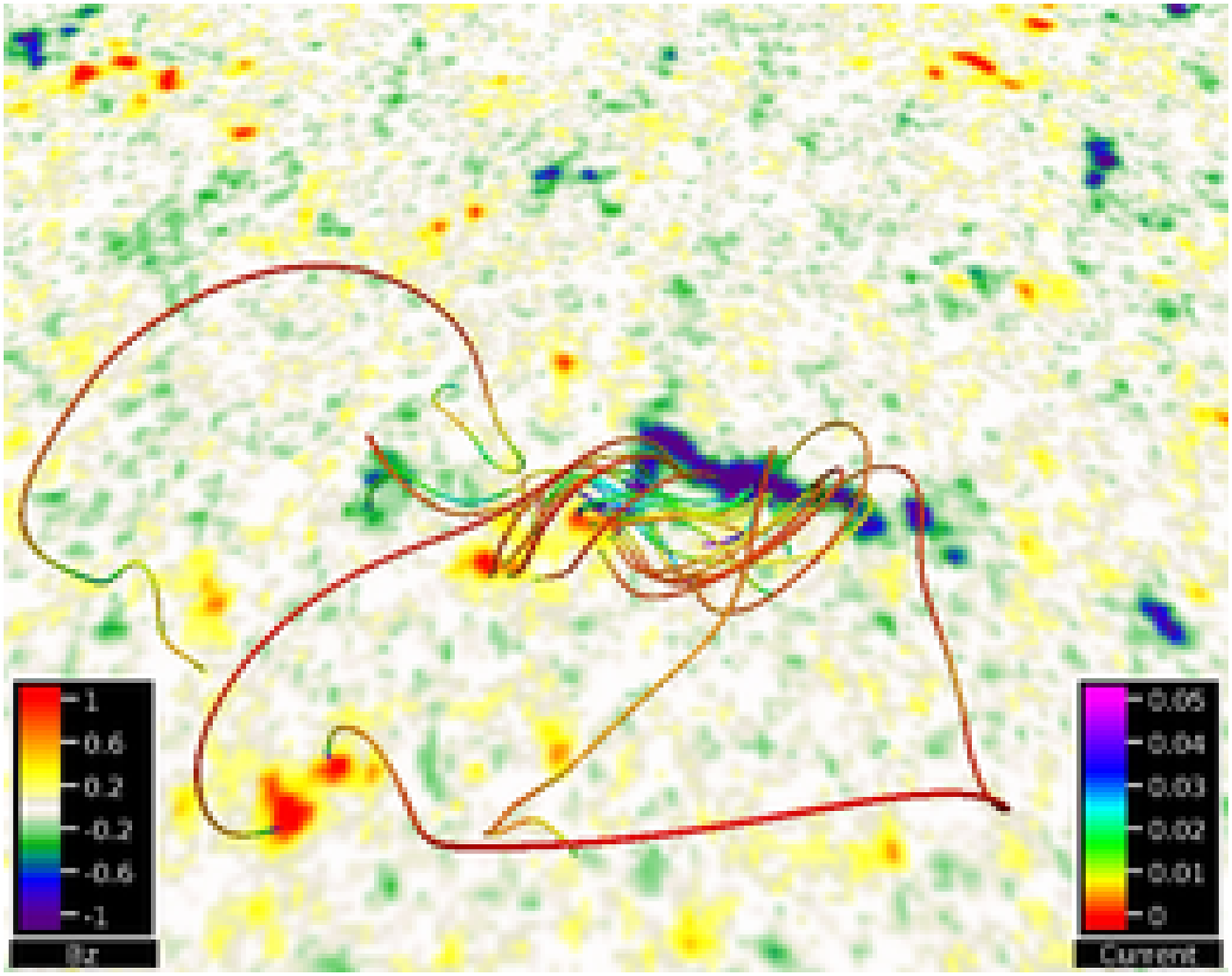}
\hfill
\includegraphics[scale=0.23]{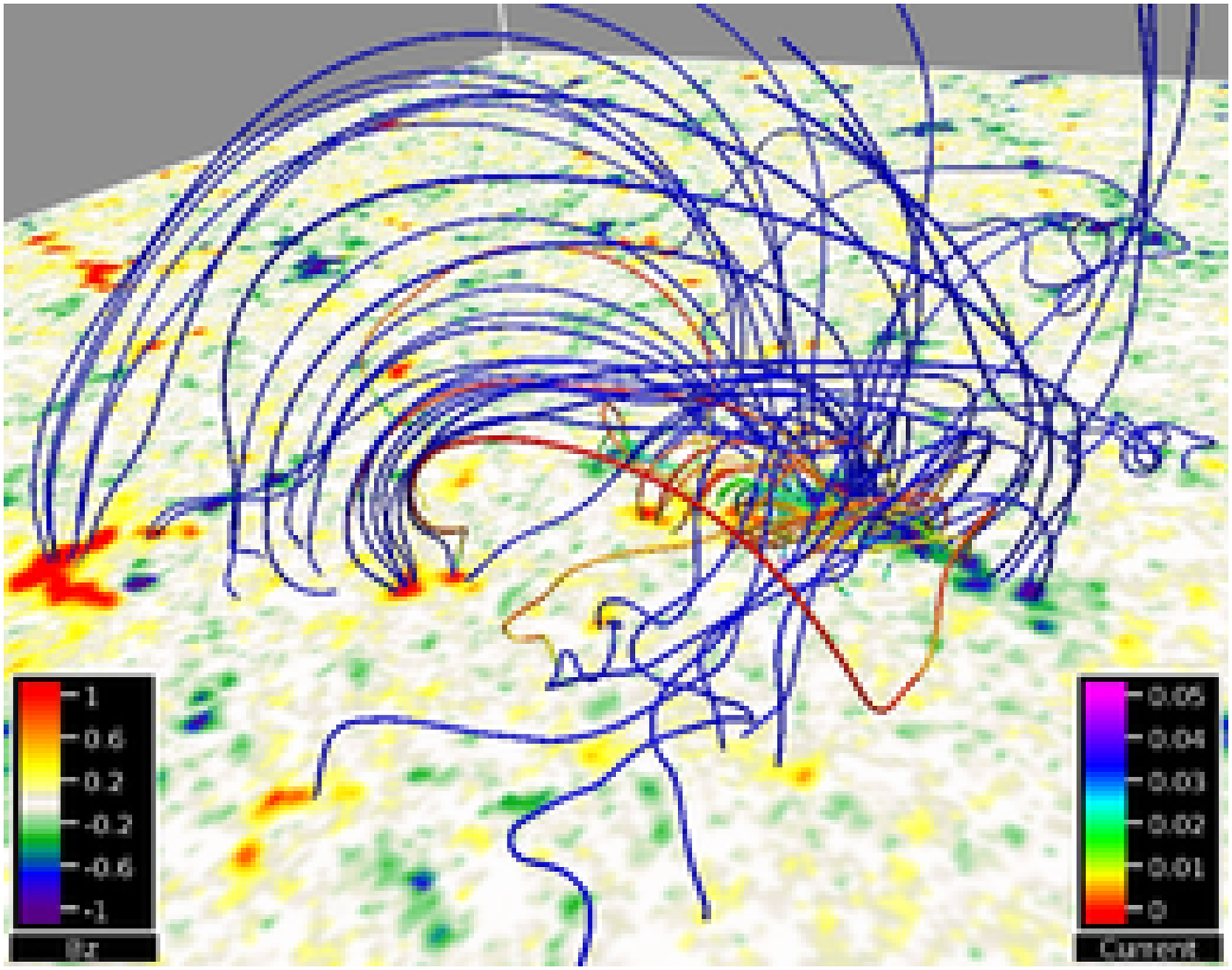}
\hfill}
\caption[]{As in Fig.~\ref{BP1_1} but for BP6. Time: 10:00 UT.}
\label{BP6} 
\end{figure*}

{\it BP6:} The local magnetic field structure is seen in the left panel of Fig.~\ref{BP6}. This shows a shared SsAS that connects the positive and negative flux regions. Also, field lines connecting from the west edge of the negative flux concentrations connect below the loop system towards the positive fluxes. These field lines form a  FR that continues deep into the arcade region, though without exiting on the opposite, eastern side of the loop system (see the movie associated with Fig.~\ref{BP6}). Some of these twisted field lines contain a strong current along a significant fraction of their length, hosting free magnetic energy that may be used in the eruption. For the large scale magnetic field structure, the emphasis is on the field in the direction of the weak expansion -- the north-east direction relative to the assumed centre of the eruption. The CBP region is embedded in a LAS that connects the negative polarity region to more distant positive fluxes in the south-east direction. Towards the north-east direction, field lines seem to form a loop system that has an east-ward alignment. 

\begin{figure*}
{\hfill
\includegraphics[scale=0.23]{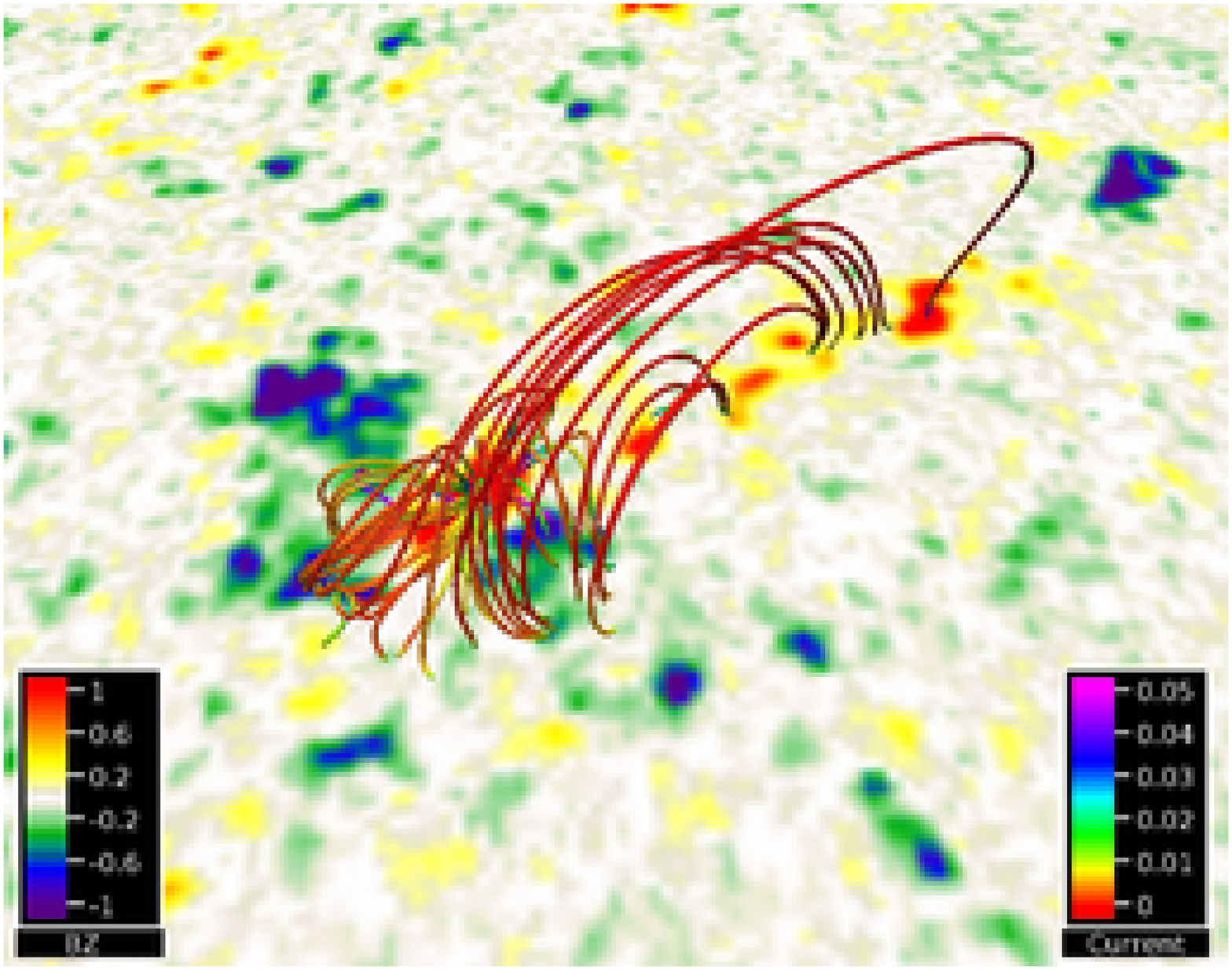}
\hfill
\includegraphics[scale=0.23]{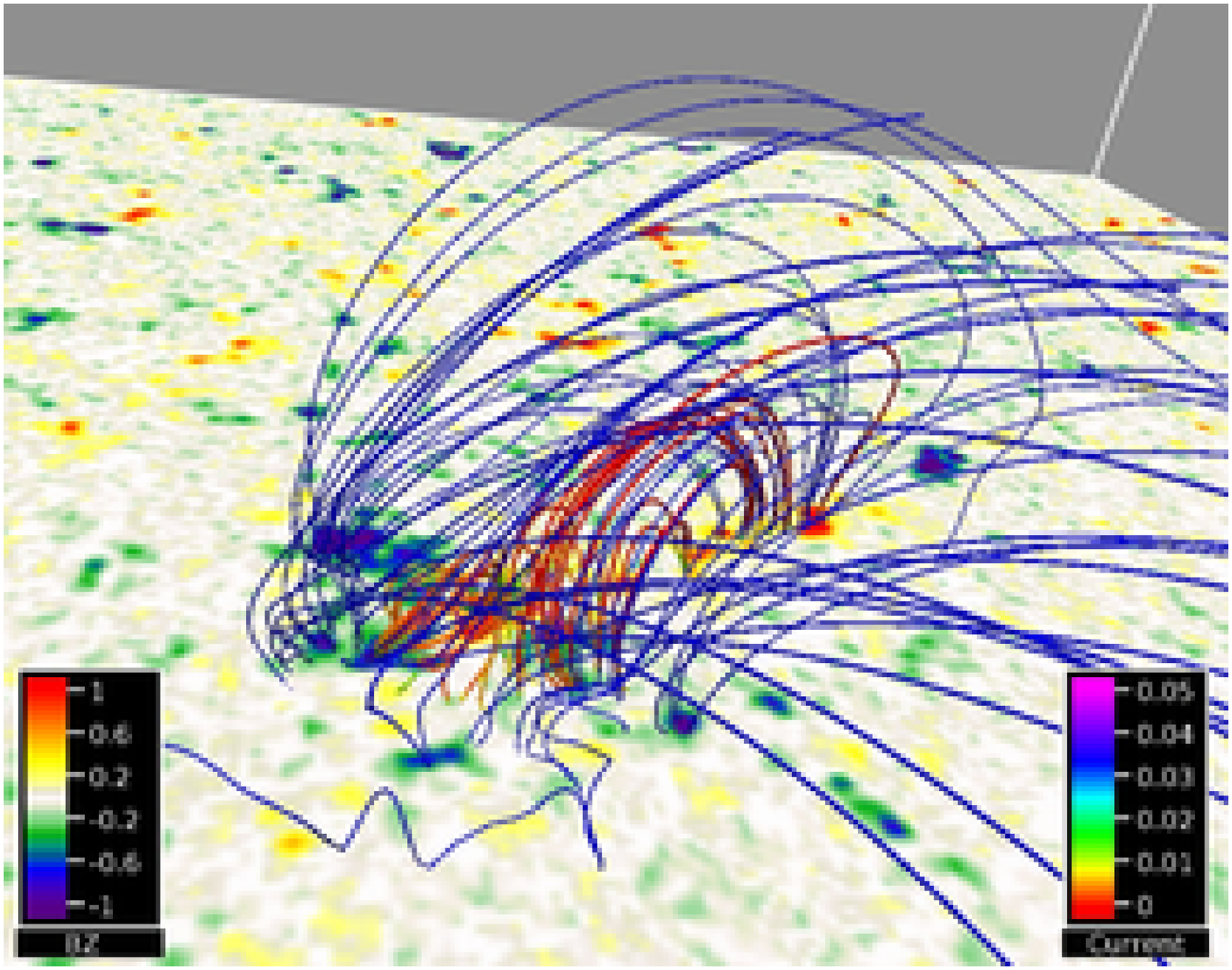}
\hfill}
\caption[]{As in Fig.~\ref{BP1_1} but for BP7--ER1.  Time: 00:18 UT.}
\label{BP7_1} 
\end{figure*}

{\it BP7--ER1:} The magnetic field line structure at the location of the eruption represents a SsAS that connects between the dominating positive polarity structure and the string of negative flux concentrations around it. This loop system shows no signatures of a  FR along its curved PIL, left panel of Fig.~\ref{BP7_1}. Part of this may be attributed to the very low grid resolution along the location of the PIL, where numerical diffusion may counter act the growth of twist (see the movie associated with Fig.~\ref{BP7_1}). Further, to obtain a twist across the PIL a shear or rotation motion of the bipolar flux structures is required, and this type of motion is difficult to capture with the present modelling approach, where only the normal component of the magnetic field is used. The large scale magnetic field defines a sequence of LASs that connects the string of negative flux concentration extending to the north-west with positive flux further to the north-west. This is a volume within which the MF expansion may take place.  


\begin{figure*}
{\hfill
\includegraphics[scale=0.23]{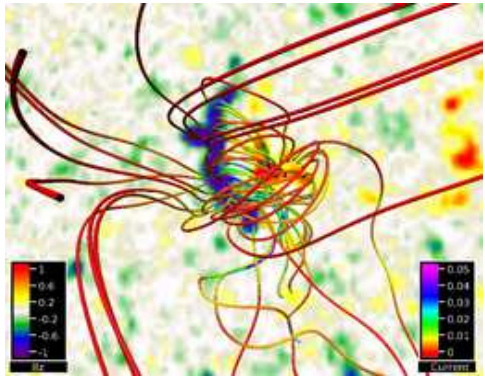}
\hfill
\includegraphics[scale=0.23]{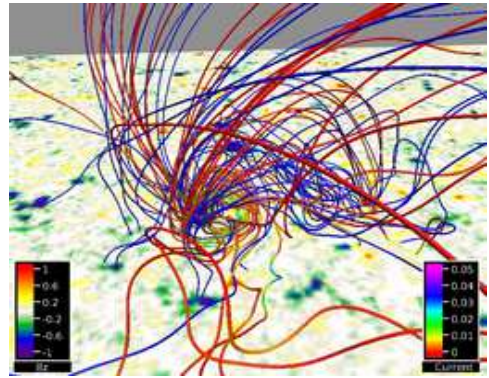}
\hfill}
\caption[]{As in Fig.~\ref{BP1_1} but for BP7--ER2.  Time: 15:18 UT.}
\label{BP7_2} 
\end{figure*}

{\it BP7--ER2:} This eruption is related to the bipole evolution of the CBP, rather than its interaction with neighbouring flux. The magnetic field configuration represents a SsAS that covers the central part of the CBP region between the dominant negative and positive flux concentrations. Part of this structure is connected by simple slightly sheared loop structures. At the west edge of the negative flux, the field lines create a  FR that connects to a distant location, left panel of Fig.~\ref{BP7_2} (see the movie associated with Fig.~\ref{BP7_2}). The eastern part of the negative magnetic flux connects to positive flux regions much further away to the west of the CBP region. This division constitutes what looks like a quasi separator surface (QSL) \citep{1995JGR...10023443P, 2007ApJ...660..863T}. The QSL allows for a fast change in the mapping in the field line connectivity, where stress may easily grow when the magnetic field is stressed by simple footpoint motions, and could be a reason for driving magnetic reconnection which could be the mechanism for the eruption in this particular case. The ejecta seen in the AIA observations is weak. It expands in the north-west direction where it soon fades into the background. This direction may be reached if the eruption is caused by the QSL close to the negative flux concentration.


\begin{figure*}
{\hfill
\includegraphics[scale=0.23]{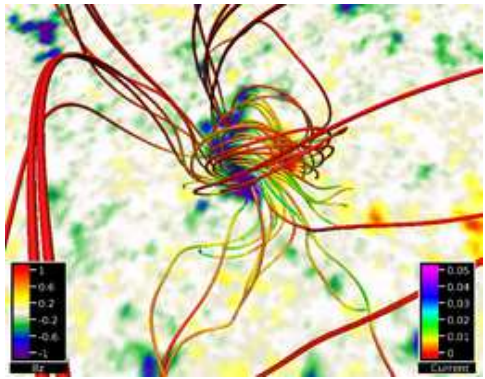}
\hfill
\includegraphics[scale=0.23]{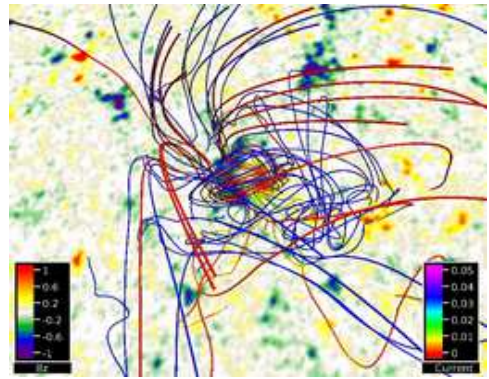}
\hfill}
\caption[]{As in Fig.~\ref{BP1_1} but for BP7--ER3. Time: 16:09 UT.}
\label{BP7_3}
\end{figure*}

{\it BP7--ER3:} The eruption takes place in the same region as the previous eruption. As the time difference between these eruptions is small, just over one hour, the magnetic structure of the region has not changed. There are some alterations to the shape of the photospheric magnetic field, with the positive flux concentration now taking up a larger part of the length of the negative polarity. This has not significantly changed the magnetic field skeleton, that is still dominated by a SsAS between the two main polarities of the CBP region, left panel of Fig.~\ref{BP7_3} (see the movie associated with Fig.~\ref{BP7_3}). From the left panel in Fig.~\ref{BP7_3} it is clear that there is a change in the shear of the magnetic field with height, a shear that has been build up since the second eruption. The large scale magnetic field is similar to that of the previous eruption, with the connectivity of the magnetic field from the negative flux concentration including the same QSL, see the right panel of Fig.~\ref{BP7_3}. 

\begin{figure*}[!h]
{\hfill
\includegraphics[scale=0.23]{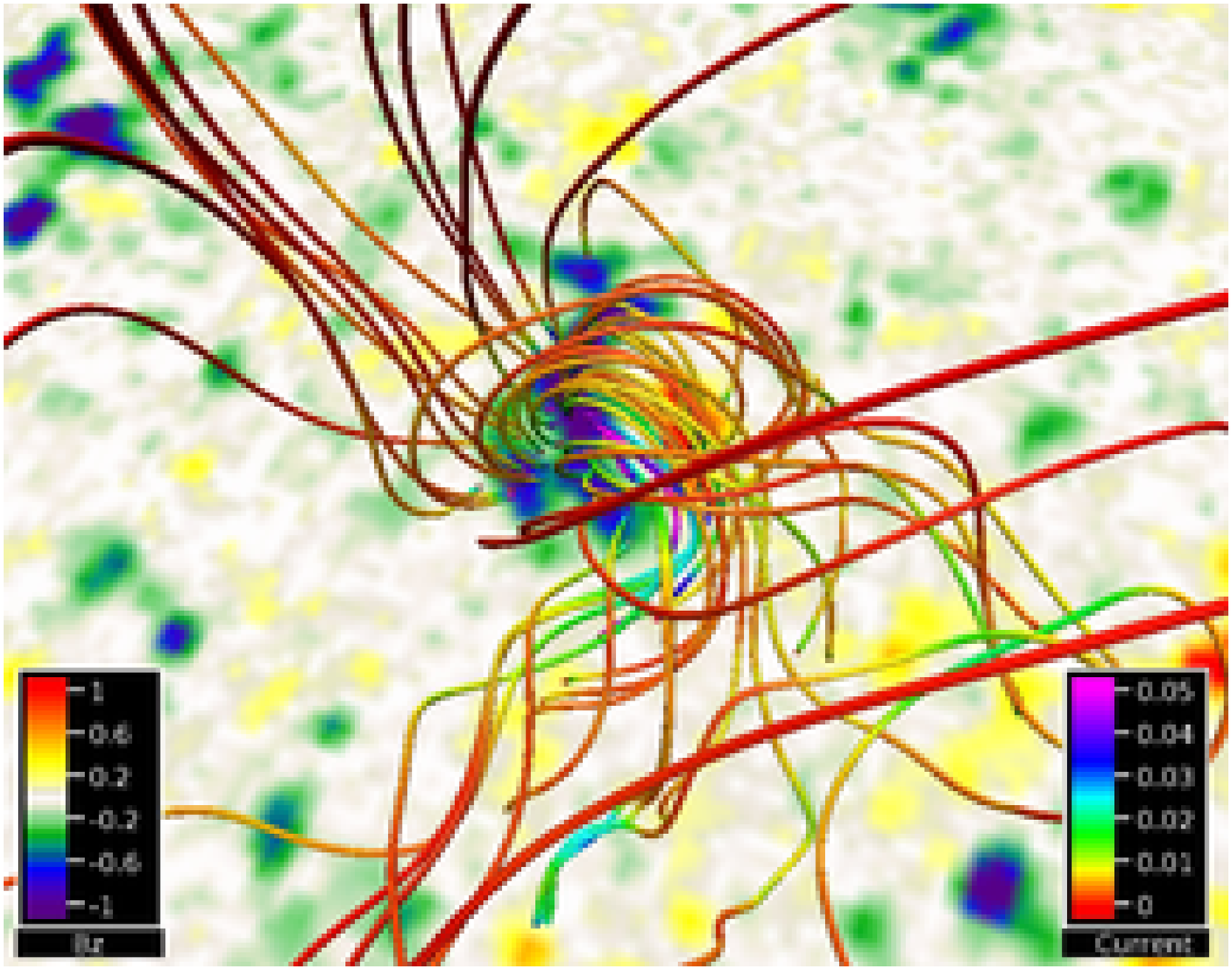}
\hfill-
\includegraphics[scale=0.23]{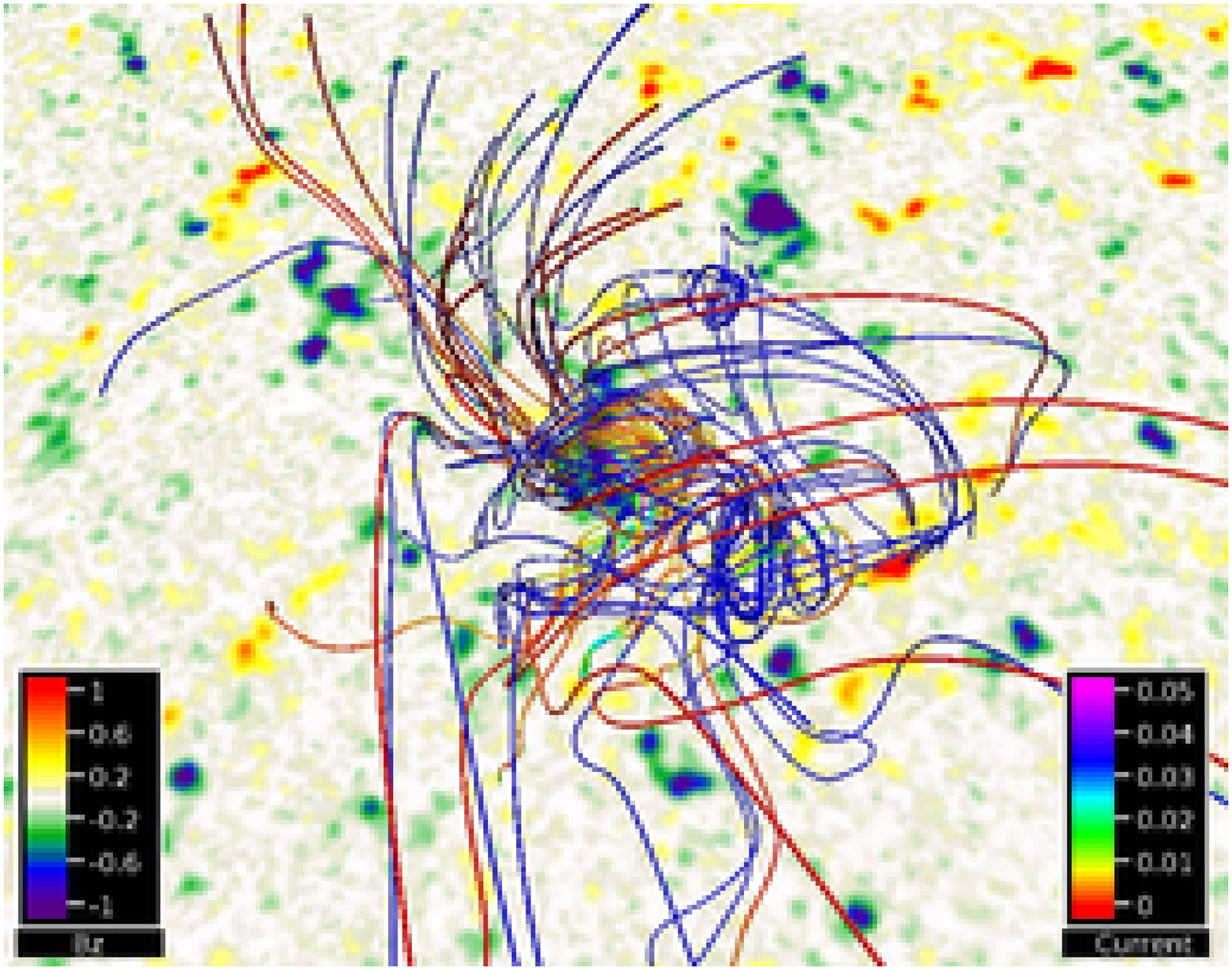}
\hfill}
\caption[]{As in Fig.~\ref{BP1_1} but for BP7--ER4. The left panel shows the field line structure in the region where the AIA emission of the eruptions initiates. The right panel shows the large scale magnetic field covering the region of the eruption. Time: 18:28 UT.}
\label{BP7_4} 
\end{figure*}


{\it BP7--ER4:} The CBP region has the same magnetic field structure as seen in the previous cases, but this time with a  FR that extends to the south of the CBP region. The magnetic field shows no clear connection towards the west, where part of the MF is seen to move, left panel of Fig.~\ref{BP7_4} (see the movie associated with Fig.~\ref{BP7_4}). The magnetic field has maintained the LAS system, but this time with a new  FR to the west of the CBP region. This structure is not involved in the dynamics seen in the AIA images.


\begin{figure*}
{\hfill
\includegraphics[scale=0.23]{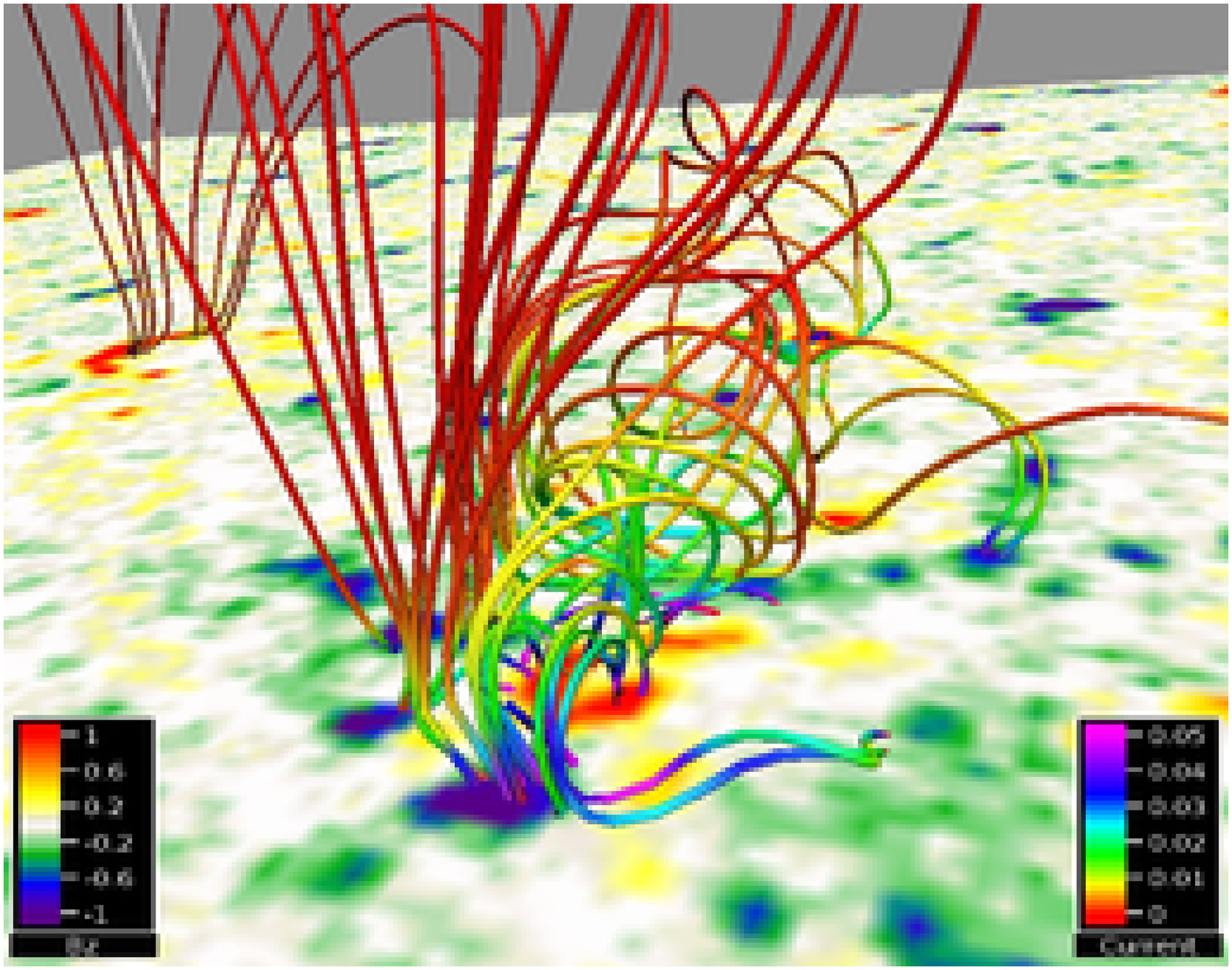}
\hfill
\includegraphics[scale=0.23]{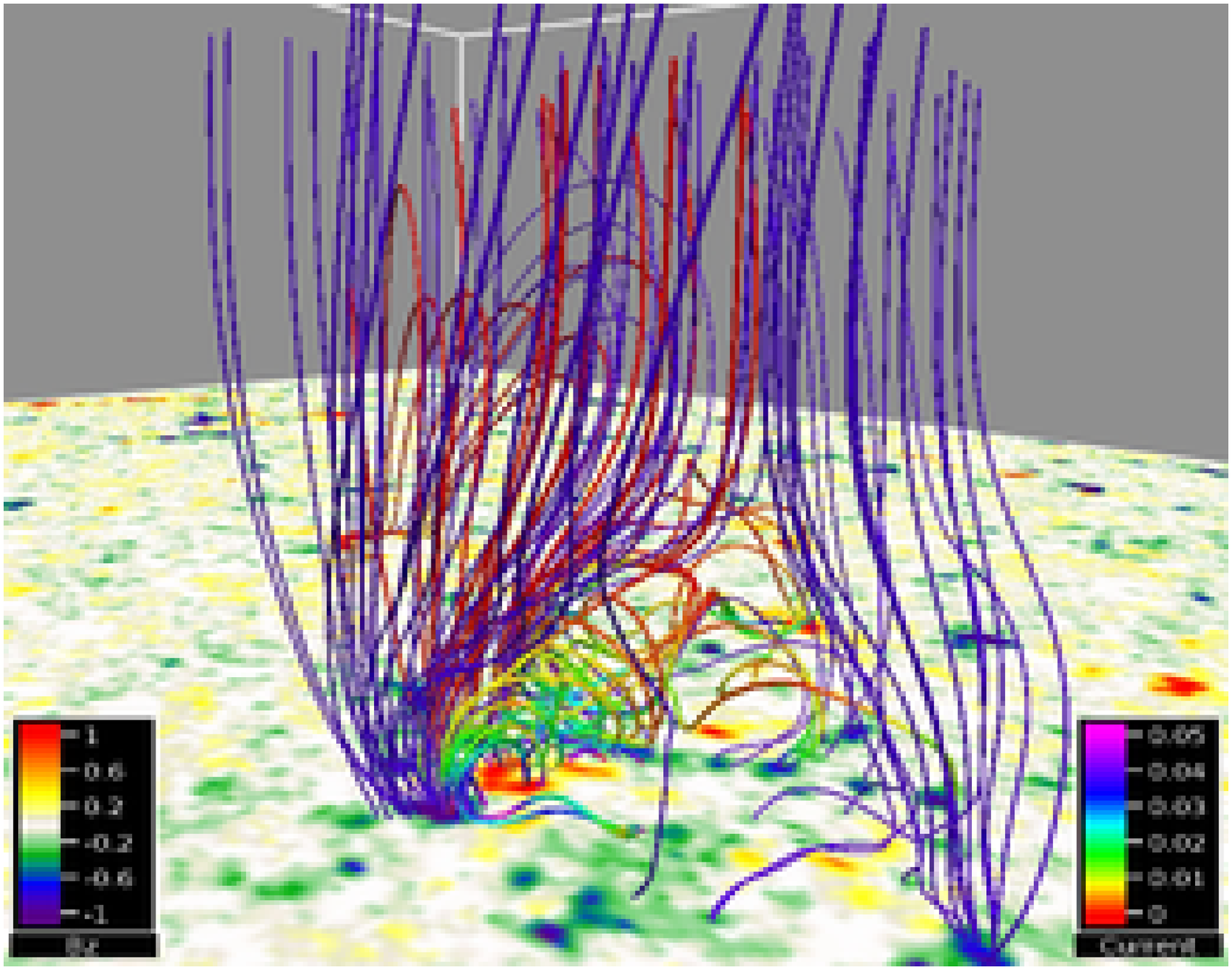}
\hfill}
\caption[]{As in Fig.~\ref{BP1_1} but for BP8--ER1. Time: 21:12 UT.}
\label{BP8_1} 
\end{figure*}

{\it BP8--ER1:} The NLFFF simulation data shows how the dominant positive polarity connects to the negative polarity south of it. The negative polarities to the east of the positive flux concentration link to a positive flux concentrated far to the east of the CBP location. This indicates the presence of a QSL between two types of field line connectivity. The QSL seems to be associated with the brightening seen in the AIA emission. This suggests that the QSL is important for the dynamical energy release, left panel of Fig.~\ref{BP8_1} (see the movie associated with Fig.~\ref{BP8_1}). On the large scale, the CBP arcade structure is embedded in what looks like an OF region defined by the negative flux concentrations on the south and east side of the dominate positive polarity, right panel of Fig.~\ref{BP8_1}. A  FR is present to the north of the positive flux region, which represents the direction of the eruption. The observed ejecta propagates in the north direction relative to the region of the positive flux patch. Magnetic field lines emanating from this part of the positive flux concentration roughly connect in this direction, but there is no obvious reason as to why the ejecta should take place from this region, as it is not directly connected to the QSL region. 

\begin{figure*}
{\hfill
\includegraphics[scale=0.23]{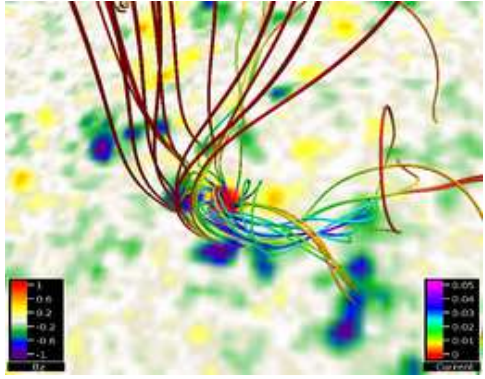}
\hfill
\includegraphics[scale=0.23]{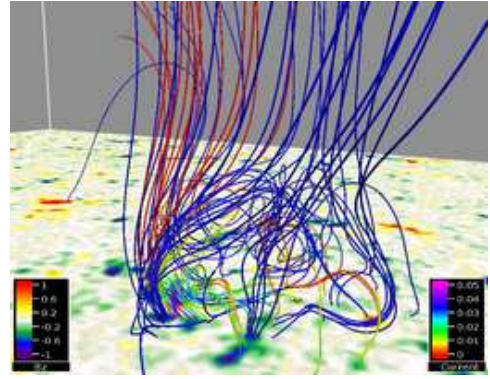}
\hfill}
\caption[]{As in Fig.~\ref{BP1_1} but for BP8--ER2.  Time: 23:42 UT.}
\label{BP8_2} 
\end{figure*}

{\it BP8--ER2:} This eruption is a large event that extends over a region of 60\arcsec\ in length. It arises from the same region as the first eruption, but expands in a north-east direction. The flux concentrations have evolved into a simpler configuration. Due to the relatively short time between the two eruptions ($\sim$2~hrs), the structure of the magnetic field in the active region has not changed significantly. In the left panel of Fig.~\ref{BP8_2} (see the movie associated with Fig.~\ref{BP8_2}) a  FR is seen to be rooted in the major positive flux concentration and it extends to the west. This structure is new, while the previous  FR located to the north has vanished. On the east-side of this flux concentration, the negative polarity flux concentrations still connect to the distant positive flux concentrations, indicating that the QSL is still present in the magnetic configuration. The large scale magnetic field structure is comparable to the first eruption, right panel of Fig.~\ref{BP8_2} (an OF). The main difference is that the eruption propagates along field lines that connect to the negative flux concentrations located to the north-east of the CBP region. The important feature is that the FR belongs to one flux system, while the propagation takes place in a different flux system. How this transfer of information from one flux region to the neighbouring flux region can take place is not clear.

\begin{figure*}
{\hfill
\includegraphics[scale=0.23]{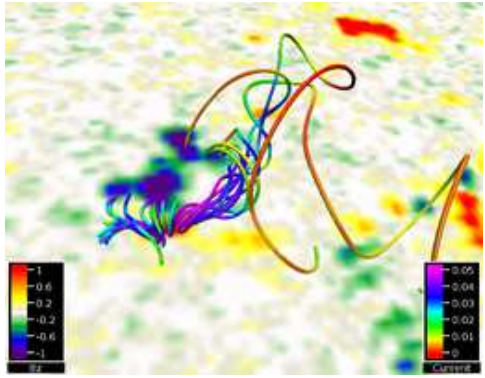}
\hfill
\includegraphics[scale=0.23]{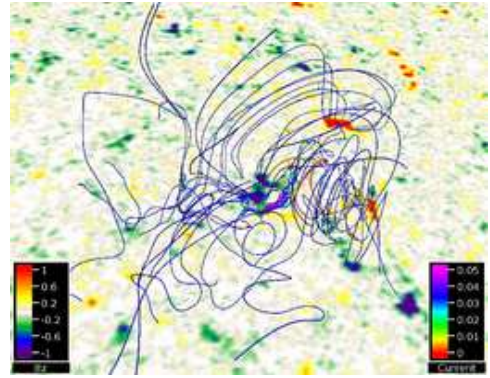}
\hfill}
\caption[]{As in Fig.~\ref{BP1_1} but for BP9.   Time: 13:33 UT.}
\label{BP9_1}
\end{figure*}

{\it BP9:} The NLFFF relaxation model has clear problems with this event. From the left panel in Fig.~\ref{BP9_1} (see the movie associated with Fig.~\ref{BP9_1}), it can be seen that the grid resolution of the magnetic flux concentrations and their PIL is very close to the pixel resolution. We may therefore only expect to see a SsAS between the minor positive polarity and the larger negative polarity. The large scale magnetic field in this region is complicated, due to the low amount of concentrated magnetic flux in the photosphere around the eruption region. This complexity is visualised in the right panel of Fig.~\ref{BP9_1}. Due to the complicated structure of the large scale magnetic field, there are no clear hints to the observed expansion of the dimming seen in the AIA channels.

\end{appendix}

\end{document}